\documentclass[twocolumn]{aastex631}
\received{\today}
\revised{\today}
\accepted{\today}
\submitjournal{ApJ}

\shorttitle{Molecular outflows in FUors with APEX}
\shortauthors{Cruz-Sáenz de Miera et al.}
\graphicspath{{./}{figures/}}
\let\tablenum\relax
\usepackage[separate-uncertainty=true,multi-part-units=single]{siunitx}
\DeclareSIUnit\parsec{pc}
\DeclareSIUnit\jansky{Jy}
\DeclareSIUnit\mas{mas}
\DeclareSIUnit\year{yr}
\newcommand{\kms}[1]{\SI[mode=text]{#1}{\kilo\meter\per\second}}
\newcommand{\ghz}[1]{\SI{#1}{\giga\hertz}}

\newcommand{\sci}[2]{\ensuremath{#1\times$10$^{#2}}}
\usepackage{cleveref}
\usepackage{booktabs}
\usepackage{upgreek}

\begin{document}
\title{An APEX study of molecular outflows in FUor-type stars}

\correspondingauthor{Fernando Cruz-S\'aenz de Miera}
\email{cruzsaenz.fernando-at-csfk.org}

\author{Fernando Cruz-S\'aenz de Miera}
\affil{Konkoly Observatory, Research Centre for Astronomy and Earth Sciences, Eötvös Loránd Research Network (ELKH),\\Konkoly-Thege Mikl\'os \'ut 15--17, 1121 Budapest, Hungary}%
\affil{CSFK, MTA Centre of Excellence, Konkoly Thege Miklós út 15--17, 1121, Budapest, Hungary}

\author{\'Agnes K\'osp\'al}
\affil{Konkoly Observatory, Research Centre for Astronomy and Earth Sciences, Eötvös Loránd Research Network (ELKH),\\Konkoly-Thege Mikl\'os \'ut 15--17, 1121 Budapest, Hungary}%
\affil{CSFK, MTA Centre of Excellence, Konkoly Thege Miklós út 15--17, 1121, Budapest, Hungary}
\affiliation{Max Planck Institute for Astronomy, Königstuhl 17, 69117 Heidelberg, Germany}%
\affiliation{ELTE E\"otv\"os Lor\'and University, Institute of Physics, P\'azm\'any P\'eter s\'et\'any 1/A, 1117 Budapest, Hungary}

\author{P\'eter \'Abrah\'am}
\affil{Konkoly Observatory, Research Centre for Astronomy and Earth Sciences, Eötvös Loránd Research Network (ELKH),\\Konkoly-Thege Mikl\'os \'ut 15--17, 1121 Budapest, Hungary}%
\affil{CSFK, MTA Centre of Excellence, Konkoly Thege Miklós út 15--17, 1121, Budapest, Hungary}
\affiliation{ELTE E\"otv\"os Lor\'and University, Institute of Physics, P\'azm\'any P\'eter s\'et\'any 1/A, 1117 Budapest, Hungary}

\author{Timea Csengeri}
\affil{Laboratoire d'astrophysique de Bordeaux, Univ. Bordeaux, CNRS, B18N, all\'ee Geoffroy Saint-Hilaire, 33615 Pessac, France}

\author{Orsolya Féher}
\affiliation{School of Physics and Astronomy, Cardiff University, Queen’s Buildings, The Parade, Cardiff CF24 3AA, UK}
\affiliation{IRAM, 300 Rue de la piscine, 38406 Saint-Martin-d’H\`eres, France}

\author{Rolf Güsten}
\affil{Max Planck Institute for Radioastronomy, Auf dem Hügel 69, 53121 Bonn, Germany}

\author{Thomas Henning}
\affil{Max Planck Institute for Astronomy, Königstuhl 17, 69117 Heidelberg, Germany}

\begin{abstract}
FU Orionis-type objects (FUors) are low-mass pre-main-sequence objects which go through a short-lived phase ($\sim$100 years) of increased mass accretion rate (from 10$^{-8}$ to 10$^{-4}$ M$_\odot$ yr$^{-1}$).
These eruptive young stars are in the early stages of stellar evolution and, thus, still deeply embedded in a massive envelope that feeds material to the circumstellar disk that is then accreted onto the star.
Some FUors drive molecular outflows, i.e.\ low-velocity wide-angle magneto-hydrodynamical winds, that inject energy and momentum back to the surrounding envelopes, and help clear the material surrounding the young star.
Here we present a $^{12}$CO (3--2), $^{13}$CO (3--2) and $^{12}$CO (4--3) survey of 20 FUor-type eruptive young stars observed with APEX\@.
We use our $^{13}$CO (3--2) observations to measure the masses of the envelopes surrounding each FUor and find an agreement with the FUor evolutionary trend found from the \SI{10}{\micro\meter} silicate feature.
We find outflows in 11 FUors, calculate their masses and other kinematic properties, and compare these with those of outflows found around quiescent young stellar objects gathered from the literature.
This comparison indicates that outflows in FUors are more massive than outflows in quiescent sources, and that FUor outflows have a higher ratio outflow mass with respect to the envelope than the quiescent sample, indicating that the eruptive young stars have lower star-forming efficiencies.
Finally, we found that the outflow forces in FUors are similar to those of quiescent young stellar objects, indicating that their accretion histories are similar or that the FUor outflows have lower velocities.
\end{abstract}

\section{Introduction}\label{sec:intro}
Jets and molecular outflows are a ubiquitous phenomenon in the process of star formation.
The former are highly collimated gas streams at high velocities ($\geq$\kms{100}), and the latter have wider opening angles and velocities between \SIlist{1;50}{\kilo\meter\per\second} in the case of low-mass stars.
Jets are detected with optical, near-infrared, radio molecular lines, and radio continuum, while the slower outflows are typically detected with molecular line tracers \citep{Frank_2014prplconf451F,Bally_2016ARAA54491B}.

Both types of mass ejection events are driven by accretion, thus the physical properties of the outflows depend on the accretion history of the star.
Indeed, evidence has shown that Class~0 objects (i.e\@. younger protostars with higher mass accretion rates) have elevated outflow mass loss rates and higher outflow forces compared to more evolved Class~I or Class~II objects \citep{Mottram_2017AA600A99M}.
The mass accretion rates from protostellar disks to protostars are expected to undergo episodic variations.
Detailed analysis of jet knots \citep[e.g.][]{Ellerbroek_2014AA563A87E,Lee_2017NatAs1E152L,Garufi_2019AA628A68G} and molecular outflow shells \citep{Plunkett_2015Natur52770P,Zhang_2019ApJ8831Z,Nony_2020AA636A38N,Vazzano_2021AA648A41V} show how the study of outflows can shed light on the accretion history of the protostars that drive them.

FU Orionis-type objects (FUors) are examples of the episodic nature of accretion \citep{Hartmann_1996ARAA34207H,Audard_2014prplconf387A,Fischer_2022PPVII}.
These eruptive young stars are low-mass protostars characterized by a sudden increase in their mass accretion rate, going from typical values of \mbox{$\sim\!10^{-8}$ M$_\odot$ yr$^{-1}$} up to \mbox{$\sim\!10^{-4}$ M$_\odot$ yr$^{-1}$}.
These events are typically detected as a $3-5$ magnitude brightening at optical and near-infrared wavelengths, and are expected to last up to a century, meaning that these events increase the final stellar mass by a significant amount.
FUor-type events generally occur in Class~I objects.
Accretion outbursts have been detected in earlier stages, e.g.\ the Class~0 HOPS~383 \citep{Safron_2015ApJ800L5S}, and in later stages, e.g.\ the Class~II Gaia20eae \citep{FCSM_2022ApJ927125C}, however, these are not considered FUors.
This differentiation is because to classify an object as a FUor, the near-infrared spectrum of the protostar must also present the spectral signatures found in the prototypical FUors \citep{Connelley_2018ApJ861145C}.
In the case a protostar shows these signatures and the photometric outburst was not detected, the source is considered \emph{FUor-like}.
And if a Class~I protostar shows an outburst and none, or a minimal amount, of the spectral signatures, then it is considered as \emph{Peculiar}.

Outflows play an important role in the star formation process as they remove angular momentum from the accretion disk, inject mass and energy into their surroundings, and clear material from the envelope \citep{Arce_2006ApJ6461070A}.
The circumstellar envelopes are the remains of the parent molecular cloud core that surround the protostar, and their properties (i.e.\,, mass and extension) are deeply connected with how evolved a young star is, with younger objects having more massive and larger envelopes than their evolved counterparts \citep{AndreMontmerle_1994ApJ420837A}.
Therefore, if the elevated accretion rates during the outbursts can inject more momentum into the envelopes via outflows, then these episodic events must play an important role in the evolution of their protostellar system.
Indeed, it is expected that after an eruption, the inner circumstellar disk becomes depleted and will be replenished by the surrounding envelope \citep{Vorobyov_2006ApJ650956V} until the system can erupt again \citep{Bell_1994ApJ427987B,Takami_2018ApJ86420T}.
Eventually, the repetitive outbursts will clear out the envelope and the young system will move to its next evolutionary phase, from Class I to Class II \citep{Green_2006ApJ6481099G,Quanz_2007ApJ668359Q,Green_2013ApJ772117G}.

Previous observations of CO rotational transitions have shown the presence of outflows in some known FUors: V1057~Cyg \citep{Rodriguez_1990PASP1021413R}, V1735~Cyg \citep{Evans_1994ApJ424793E}, L1551~IRS~5 \citep[and references therein]{Wu_2009ApJ698184W}, V883~Ori \citep{RuizRodriguez_2017MNRAS4683266R}, Reipurth~50 \citep{RuizRodriguez_2017MNRAS4663519R}, FU~Ori \citep{Hales_2015ApJ812134H}, V1647~Ori \citep{Principe_2018MNRAS473879P}, V2775~Ori \citep{Zurlo_2017MNRAS465834Z}, V346~Nor \citep{Kospal_2017ApJ84345K} and V900~Mon \citep{Takami_2019ApJ884146T}.
In other cases, optical and near-infrared spectroscopy have shown indication of high-velocity jets: Z~CMa \citep{Poetzel_1989AA224L13P}, V899~Mon \citep{Ninan_2015ApJ8154N}, iPTF~15afq \citep{Hillenbrand_2019ATel133211H}, and V346~Nor \citep{Kospal_2020ApJ889148K}.

Each of the aforementioned studies focused on a single FUor-type object or on a few of them, preventing a statistical analysis of their properties.
In this paper we present a systematic study of the envelopes surrounding FUors, and we search for outflows among our full sample.
We then compare our results with outflows found in YSOs that are currently quiescent and for which it is unknown whether they experienced an outburst or not.
As outflows found at thousands of astronomical units are an indication of the accretion history of a protostar, this comparison allows us to examine how comparable is the histories of the FUors with those of the quiescent sample.
The triggering mechanisms behind the FUor-type outbursts is still not understood, however, it is possible that an examination of the differences between the two samples might hint that FUors are protostar with intrinsic differences that caused the outburst.
Alternatively, it could show the samples are similar and, thus, we cannot rule out that quiescent sources experienced FUor-type outbursts in the past.

The structure of the paper is as follows.
The observed sample is briefly introduced in \autoref{sec:sample}, while in \autoref{sec:observations} we describe the observations and the data reduction.
In \autoref{sec:results} we present the distribution of the gas in the environment surrounding the FUors, and the properties of the integrated line profiles.
The main goal of this paper is to study the properties of the circumstellar gas, this analysis is found in \autoref{sec:analysis}, including the characterization of the molecular outflows where detected.
In \autoref{sec:discussion} we compare the outflows found in the FUor sample with non-outbursting sources and draw our conclusions.
Finally, in \autoref{sec:theend} we summarize our work and present our main findings.

\section{Sample}\label{sec:sample}
Our sample is composed of 20 eruptive young stars, including most of the known FUors accessible from the APEX site \citep{Audard_2014prplconf387A,Connelley_2018ApJ861145C}.
We note that not all the targets in our list are considered FUors as some objects are cataloged as \emph{FUor-like} objects.
This subclassification is used when the photometric outburst was not detected but their near-infrared spectrum shows features similar to those of the prototypical FUors \citep[e.g.\ BBW 76,][]{Connelley_2018ApJ861145C}.
The target list also includes eruptive young stars with peculiar accretion histories (e.g.\ V1647~Ori) and a massive star with a powerful accretion outburst (V723~Car).
The non-FUor objects were included because of their sudden increases of their mass accretion rate, and thus the properties of their outflows could be affected.
The first part of the sample, composed of eight targets, was analyzed by \citet{Kospal_2017ApJ836226K}, where they found outflows in three objects: HBC~494, Haro~5a~IRS and V346~Nor.
Here we will analyze the full sample, including a re-processing of the target list presented in \citet{Kospal_2017ApJ836226K}.
The sample presented in this paper includes $\sim$50\% of the currently known FUors and FUor-like objects \citep{Connelley_2018ApJ861145C}.
The full target list is presented in \autoref{tab:values}.

\begin{deluxetable*}{ccccccc}
\tablecaption{Objects observed.\label{tab:values}}
\tablehead{
\colhead{Name} & \colhead{Coordinates} & \colhead{Class\tablenotemark{a}} & \colhead{FUor classification} & \colhead{$v_\mathrm{LSR}$\tablenotemark{b}} & \colhead{Distance\tablenotemark{c}} & \colhead{$L_\mathrm{bol}$\tablenotemark{d}}\\
\colhead{} & \colhead{} & \colhead{} & \colhead{} & \colhead{[\si{\kilo\meter\per\second}]} & \colhead{[\si{\parsec}]} & \colhead{[L$_\odot$]}
}
\startdata
L1551~IRS~5                  & 04:31:34.07 +18:08:04.9   & I    & FUor-like      & 6.46     & 147  & 25  \\
V582~Aur                     & 05:25:51.97 +34:52:30.0   & F    & Bona fide FUor & $-$10.85 & 1320 & 146 \\
Haro~5a~IRS                  & 05:35:26.75 $-$05:03:55.1 & I    & FUor-like      & 10.90    & 391  & 50  \\
V883~Ori                     & 05:38:18.09 $-$07:02:25.9 & F    & Bona fide FUor & 4.10     & 388  & 400 \\
Reipurth~50\tablenotemark{e} & 05:40:27.45 $-$07:27:30.0 & I    & Peculiar       & 3.76     & 460  & 300 \\
FU~Ori                       & 05:45:22.36 +09:04:12.2   & I/II & Bona fide FUor & 11.96    & 402  & 420 \\
V1647~Ori                    & 05:46:13.13 $-$00:06:04.8 & I/II & Peculiar       & 10.06    & 388  & 39  \\
V2775~Ori                    & 05:42:48.48 $-$08:16:34.7 & I    & Bona fide Fuor & 3.08     & 428  & 25  \\
V899~Mon                     & 06:09:19.24 $-$06:41:55.8 & F/II & Peculiar       & 9.57     & 785  & 419 \\
AR~6A\tablenotemark{f}       & 06:40:59.30 +09:35:52.3   & I    & Peculiar       & 5.02     & 890  & 450 \\
V900~Mon                     & 06:57:22.22 $-$08:23:17.6 & I    & Bona fide FUor & 13.77    & 1130 & 106 \\
V960~Mon                     & 06:59:31.58 $-$04:05:27.7 & II   & Bona fide FUor & 23.81    & 2068 & \\
Z~CMa                        & 07:03:43.15 $-$11:33:06.2 & I    & FUor-like      & 13.91    & 1150 & 500 \\
iPTF~15afq\tablenotemark{g}  & 07:09:21.39 $-$10:29:34.4 & I/F  & Peculiar       & 14.04    & 920  & 7.2 \\
BBW~76\tablenotemark{e}      & 07:50:35.59 $-$33:06:23.9 & I    & FUor-like      & 17.64    & 1040 & 287 \\
V723~Car                     & 10:43:23.44 $-$59:33:55.3 & I    & Peculiar       & $-$19.58 & 2500 & 4000\\
GM~Cha                       & 11:09:28.55 $-$76:33:28.1 & I/II & Peculiar       & 4.86     & 160  & 1.5 \\
V346~Nor                     & 16:32:32.19 $-$44:55:30.7 & 0/I  & Peculiar       & $-$3.08  & 700  & 135 \\
OO~Ser                       & 18:29:49.13 +01:16:20.6   & I    & Peculiar       & 8.36     & 311  & 31  \\
HBC~687\tablenotemark{h}     & 19:29:00.87 +09:38:42.9   & II   & FUor-like      & 16.98    & 400  & 10  \\
\enddata
\tablenotetext{a}{Here we refer as Class to the classification based on the shape of its SED \citep{Lada_1987IAUS1151L,Andre_1993ApJ406122A,Greene_1994ApJ434614G}.}
\tablenotetext{b}{See \autoref{ss:sysvel}.}
\tablenotetext{c}{See \autoref{ss:distances}.}
\tablenotetext{d}{Obtained from the literature \citep[e.g.][and references therein]{Audard_2014prplconf387A,Connelley_2018ApJ861145C}.}
\tablenotetext{e}{Reipurth~50 and BBW~76 are labeled as HBC~494 and Bran~76 in \citet{Kospal_2017ApJ836226K}.}
\tablenotetext{f}{Also known as V912~Mon.}
\tablenotetext{g}{Also known as Gaia19fct.}
\tablenotetext{h}{Also known as Parsamian~21.}
\end{deluxetable*}

\section{Observations and data reduction}\label{sec:observations}
We carried out two programs with the FLASH$^+$ receiver \citep{Klein_2014ITTST4588K} at the APEX telescope \citep{Gusten_2006SPIE6267E14G} to measure the $^{12}$CO~(3--2), $^{13}$CO~(3--2), and $^{12}$CO~(4--3) lines towards our targets.
Program 094.F-9508 was observed between 2014 August 23--28 and program 098.F-9505 between 2016 August 25 and 2016 September 10.
Both programs used the same technical setup and reduction process.
The lower frequency channel was tuned to \ghz{344.2} in USB to cover the $^{13}$CO (3--2) at \ghz{330.588}, and the $^{12}$CO (3--2) at \ghz{345.796}, respectively.
The higher frequency channel was tuned to the $^{12}$CO (4--3) line at \ghz{461.041} in USB\@.
We used the XFFTS backends providing a nominal \SI{38}{\kilo\hertz} spectral resolution for the $J$ = 3--2 lines and \SI{76}{\kilo\hertz} for the $J$ = 4--3 line, these resulted in spectral resolutions of $\sim$\SI{34.5}{\meter\per\second}, $\sim$\SI{32.9}{\meter\per\second} and $\sim$\SI{49.4}{\meter\per\second} for the $^{13}$CO (3--2), CO (3--2) and CO (4--3) lines, respectively..
For each target, \SI{120}{\arcsecond}${\times}$\SI{120}{\arcsecond} on-the-fly (OTF) maps were obtained at \SI{6}{\arcsecond\per\second}, using a relative reference off position \SI{1000}{\arcsecond} away in right ascension.

We removed a first order baseline from the spectra, and calibrated the data using a main beam efficiency of 0.73 and 0.60 at \SIlist{352;464}{\giga\hertz}, respectively, and the values were converted to Jansky using \SIlist{41;48}{\jansky\per\kelvin} at \SIlist{352;464}{\giga\hertz}, respectively.
We calculated the noise levels of each CO line by first selecting the first and the last 100 channels of each cube (individually confirmed to be free of line emission), calculated the noise levels for each FUor using these channels, and then we calculated the median noise level of all FUors to obtain representative values.
The rms noise levels, at the native spectral resolution mentioned earlier, are \SI{3.6}{\jansky} for $^{13}$CO~(3--2), \SI{3.7}{\jansky} for $^{12}$CO~(3--2), and \SI{9.4}{\jansky} for $^{12}$CO~(4--3).
The telescope's half-power beam-width is $19$\farcs$2$, and $15$\farcs$3$ at the corresponding frequencies.
As mentioned earlier, the first half of the survey has already been published by \citet{Kospal_2017ApJ836226K}, and here we use their calibrated data for our analyses.

\section{Results}\label{sec:results}
\subsection{Distribution of gas}\label{ss:moment0}
\begin{figure*}
\centering
\includegraphics[width=\textwidth]{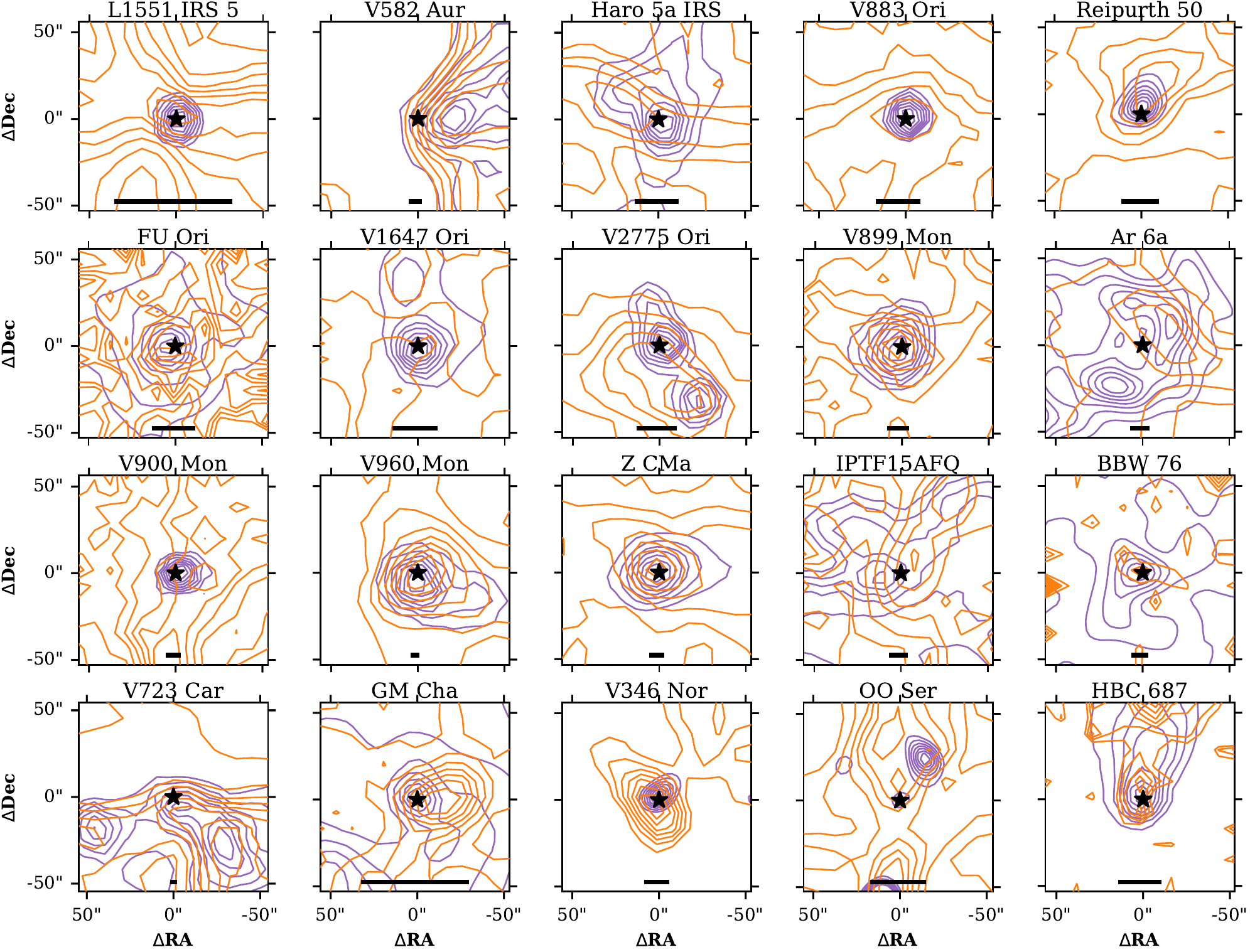}
  \caption{Integrated intensity (Moment 0) maps of our targets for the $^{12}$CO~(3--2) line observed with APEX (orange contours). The Moment 0 maps were generated by integrating the full spectral cube in order to produce an unbiased map. The purple contours are the \SI{250}{\micro\meter} continuum emission from \emph{Herschel} for most of our targets, the two exceptions are V900~Mon and Z~CMa where we show contours of the \SI{850}{\micro\meter} continuum emission from the JCMT (see \autoref{ss:moment0}). The CO and the dust contours are plotted with levels at 0.3, 0.4, \ldots, 0.9 of the peak intensity, and are meant to be representative. The star symbols indicate the nominal positions of the protostars. The bars at the bottom of each panel represent \SI{10000}{\astronomicalunit}.\label{fig:moment0}}
\end{figure*}

We constructed velocity integrated emission maps (Moment 0) for $^{12}$CO (3--2) using all channels in our data cubes.
The resulting maps are presented in \autoref{fig:moment0}.
Some young eruptive stars are still deeply embedded, therefore, it is possible that the observed CO emission originates from the remaining material in their surrounding envelope.
In order to verify that our CO detections come from the FUors, we compared our Moment~0 maps with the dust continuum emission.
We searched for \SI{250}{\micro\meter} continuum maps taken with \emph{Herschel}/SPIRE maps in the \emph{Herschel} Science Archive\footnote{\url{http://archives.esac.esa.int/hsa/whsa/}} and found data at an angular resolution of 17.6$''$ for 18 sources.
For the two remaining sources (V900~Mon and Z~CMa), we searched the Canadian Astronomy Data Centre\footnote{\url{https://www.cadc-ccda.hia-iha.nrc-cnrc.gc.ca/en/}} for archival \SI{850}{\micro\meter} observations taken with the James Clerk Maxwell Telescope\footnote{The James Clerk Maxwell Telescope has historically been operated by the Joint Astronomy Centre on behalf of the Science and Technology Facilities Council of the United Kingdom, the National Research Council of Canada and the Netherlands Organisation for Scientific Research.} (JCMT) with an angular resolution of 14.5$''$.
In the cases of L1551~IRS~5, Haro~5a~IRS, V883~Ori, Reipurth~50, V899~Mon, V960~Mon, Z~CMa, V346~Nor, GM~Cha, and HBC~687, the peaks of both the dust emission and the gas emission are located at the position of the protostar.
For five of our targets (V582~Aur, AR~6A, iPTF~15afq, V723~Car, and OO~Ser) the brighter peaks of both gas and dust emission are offset from the position of the protostar.
The continuum peaks to these five sources are 27$''$ to the Southeast, 12$''$ to the East, 28$''$ to the Northwest, 22$''$ to the West, and 6$''$ to the Southwest, respectively.
In the case of OO~Ser, there is continuum emission toward the position of the FUor, however the brightest peak is the one previously mentioned.
We find that for FU~Ori, V1647~Ori, V2775~Ori, and V900~Mon, the dust emission is located at the position of the protostar, however, the peak of the gas emission is offset.
Finally, in the case of BBW~76, the dust emission peaks at the position of the source, however, the CO map shows that the emission is weak and extended.

\subsection{Systemic velocity}\label{ss:sysvel}
\begin{figure*}
\centering
\includegraphics[width=\textwidth]{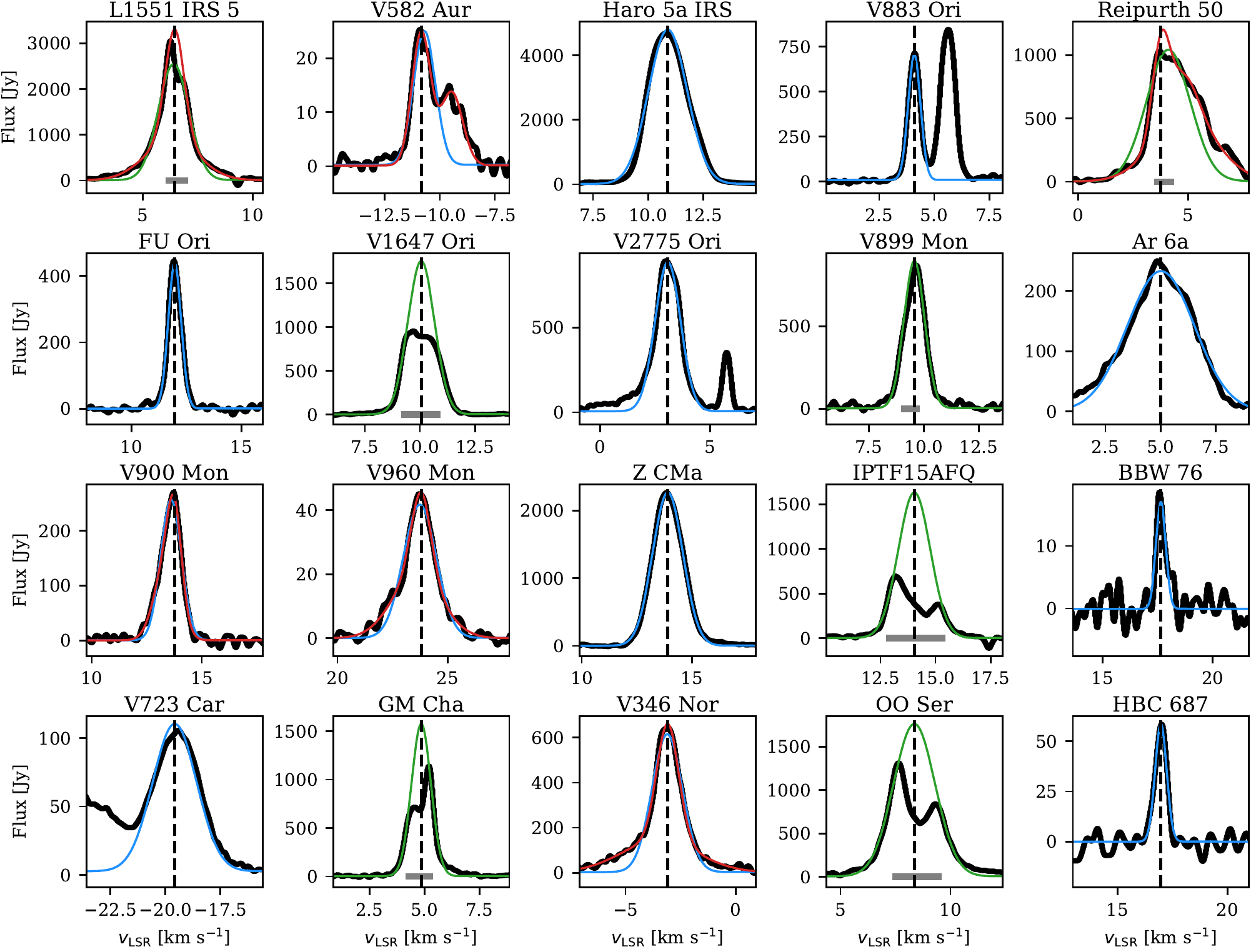}
  \caption{Line profiles of $^{13}$CO (black lines) extracted using a circular aperture with radius of \SI{10000}{\astronomicalunit} to determine the systemic velocities. The blue lines indicate the best-fit Gaussian when using the full velocity range, and the green lines where the fit was done without including velocities close to the peak. The gray horizontal line at \SI{0}{\jansky} indicates the range of velocities excluded from this second fit. The red lines show the best-fit when using two or three Gaussians. The vertical dashed line indicates the systemic velocity of each FUor. In the cases of V883~Ori, V2775~Ori and V723~Car, i.e.\ FUors with known emission from other sources in the same line of sight, we did not use additional Gaussians to fit the additional components because they can be easily separate from the single Gaussian fit. See \autoref{ss:sysvel} for details.\label{fig:sysvel}}
\end{figure*}

As it is shown below, to estimate the kinematic properties of the outflows, we need a reliable estimate of the systemic velocity for each target so that we can measure the velocity of the outflow relative to the protostar.
The systemic velocities of our targets were estimated by fitting a Gaussian function to the line profile of\ $^{13}$CO, extracted using a circular aperture with radius of \SI{10000}{\astronomicalunit}, and using the center of the best-fitting Gaussian as the systemic velocity.
Due to the proximity of L1551~IRS~5 and the field-of-view of our observations, we had to use a smaller aperture of \SI{8000}{\astronomicalunit} for this FUor.
The line profiles and the best-fit Gaussians are shown in \autoref{fig:sysvel}.

A number of sources have complicated line profiles that could not be fitted by a single Gaussian.
Some of these show asymmetric dips around the peak of the emission (e.g.\ V1647~Ori, GM~Cha, and OO~Ser), which can be due to the self-absorption of the envelope or due to the rotation of the gas.
The former scenario is more likely based on the inspection of the channel maps of $^{13}$CO, thus, for these sources, we discarded the velocity range of the dip and fitted the Gaussian function using the remaining velocities.
The line profiles of other objects show asymmetric shapes out to the wings of the line profiles (e.g.\ V582~Aur, Reipurth~50, and V346~Nor), an indication of multiple components (e.g.\ envelope, outflows, Keplerian disk, or unrelated gas in the same line-of-sight) showing emission at $^{13}$CO\@.
For these objects, we fitted a combination of two or three Gaussian functions to the line profile, and used the best-fit mean of the Gaussian with the highest amplitude to estimate the systemic velocity.

To verify our estimated systemic velocities, we searched for the velocity around which the line profile is most symmetrical.
As expected, we found that the sources with asymmetrical line profiles have the largest differences, but still less than \kms{1}.
For the symmetrical sources, the differences are less than \kms{0.3}.
In addition, we examined the channel maps of each target to confirm our systemic velocity estimates.

As a final step, we compared our estimates with those from the literature.
The differences between our estimates and those obtained from previous observations are
\kms{0.04} for L1551~IRS~5 \citep{Wu_2009ApJ698184W},
\kms{0.02} for V2775~Ori \citep{Zurlo_2017MNRAS465834Z},
\kms{0.04} for GM~Cha \citep{Mottram_2017AA600A99M},
\kms{0.20} for V883~Ori \citep{RuizRodriguez_2017MNRAS4683266R},
\kms{0.06} for V1647~Ori \citep{Principe_2018MNRAS473879P},
\kms{0.35} for V582~Aur \citep{Abraham_2018ApJ85328A},
\kms{0.27} for V900~Mon \citep{Takami_2019ApJ884146T},
\kms{0.56} for FU~Ori~North \citep[][who resolved the binary system with an angular resolution of 0.05$"$ using ALMA]{Perez_2020ApJ88959P}.
In the cases of Haro~5a~IRS, AR~6A, BBW~76, OO~Ser, and HBC~687 the estimates by \citep{Kospal_2017ApJ836226K} are in agreement within \kms{0.30}.
For V899~Mon, V960~Mon, Z~CMa, iPTF~15afq, and V723~Car these are the first estimates of their systemic velocity.
Two of our measurements deviate from those determined by interferometric observations of C$^{18}$O\@: V346~Nor and Reipurth~50.
In the case of the former, \citet{Kospal_2017ApJ84345K} found the line profile peaks at \kms{-3.55}, indicating a difference of \kms{0.47} from our estimate, and in the case of the latter FUor, \citet{RuizRodriguez_2017MNRAS4663519R} determined a systemic velocity of \kms{4.6}, a value \kms{0.77} different from ours.
It is likely that the larger differences are due to the interferometric observations resolving out emission from the extended envelopes.
The final values for the systemic velocities are presented in \autoref{tab:values}.

\subsection{Line Profiles}\label{ss:line_profiles}
\begin{figure*}
\centering
\includegraphics[width=\linewidth]{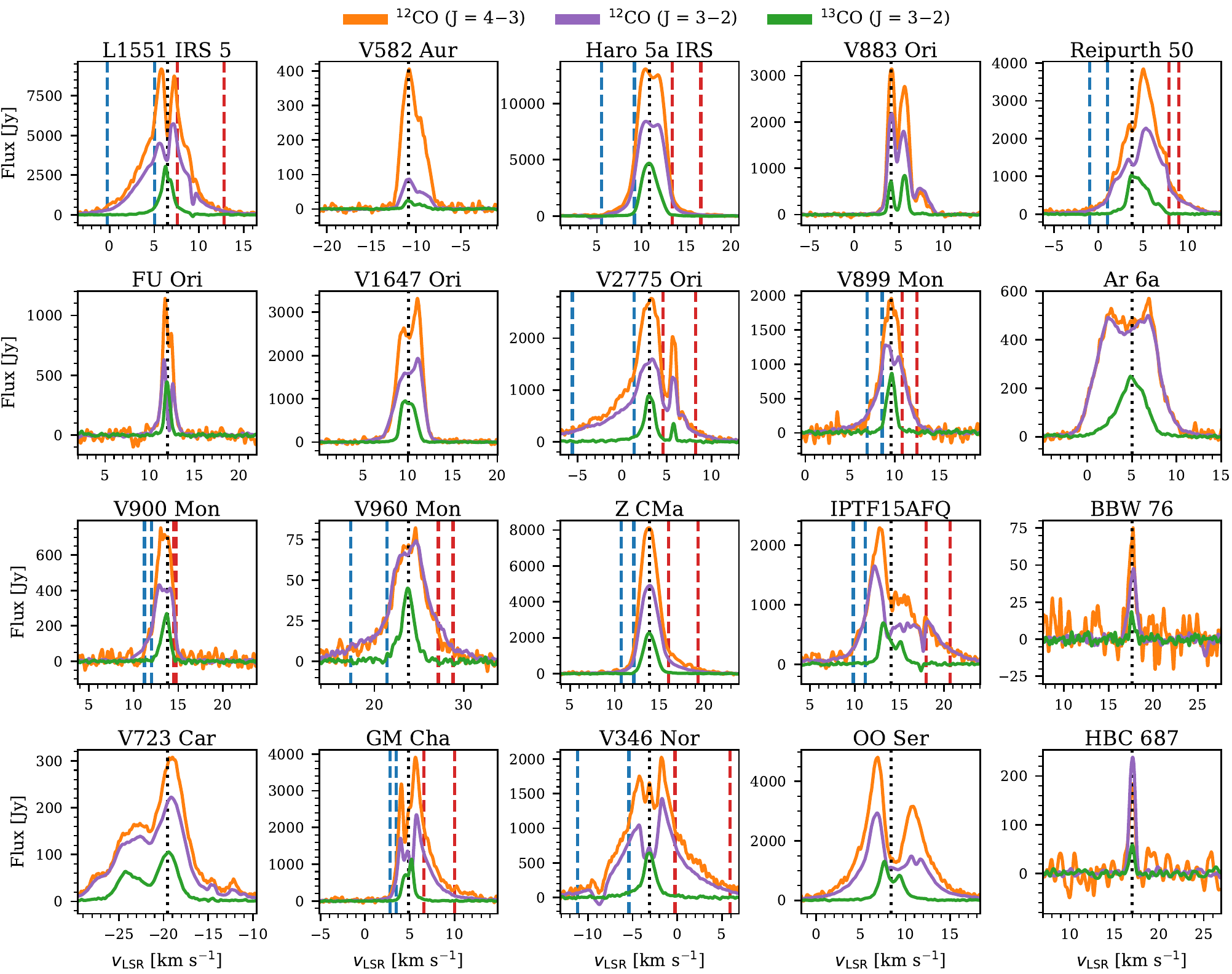}
  \caption{CO line profiles of our targets observed with APEX\@. The vertical dotted line is the systemic velocity. The vertical dashed lines are the range of velocities of the CO (3--2) outflows. The line profiles have been smoothed for presentation purposes.\label{fig:spectra}}
\end{figure*}

In order to examine the outflows using their line profiles, we must select apertures that cover the gas emission.
We began by exploring the channel maps of the two $^{12}$CO transitions and checking which channels and which regions show emission above the 3$\sigma$ contour level.
The channels with wide extended emission that showed little variations from channel to channel were considered as envelope emission.
Then we inspected the blue- and red-shifted channel maps for emission similar to what is found in outflows, i.e.\ emission whose red-shifted channels is in the opposite side from the blue-shifted channels with respect to the expected position of the star, and emission that is generally more extended in the channels with velocities closer to the systemic velocity and more compact towards higher velocities.
Finally, we created a polygon whose shape would cover this emission in both transitions.
For the targets where the CO emission does not the morphology described above, the spectra were extracted using a \SI[mode=text]{10000}{\astronomicalunit} aperture centered on the nominal position of the protostar.
The only exception is L1551~IRS~5, where we used a circular aperture with a radius of \SI[mode=text]{8000}{\astronomicalunit}, due to the proximity of this source (see below) and the size of our CO map.
For each target, we used the same aperture in the three CO maps.
The aperture used for each target can be seen in their channel maps in \autoref{app:channelmaps}, and the spectral line profiles integrated over these apertures for all three observed CO lines are presented in \autoref{fig:spectra}.

The line profiles and the channel maps show contamination caused by faint extended emission in four of the FUors: V582~Aur (at $\sim$\kms{-9}), V883~Ori (at $\sim$\kms{5.5}), V2775~Ori (at $\sim$\kms{5.9}), and V723~Car ($\sim$\kms{-24}).
The peaks in the V883~Ori profiles were reported by \citet{White_2019ApJ87721W}, and the blueshifted broad feature in V582~Aur was discussed by \citet{Abraham_2018ApJ85328A}.

HBC~687, BBW~76, FU~Ori, and V883~Ori show the narrowest lines in our sample.
The line profiles of V1647~Ori, V900~Mon, and Z~CMa are slightly wider and do not show obvious indications of wings caused by high velocity outflows.
The remainder of the sources exhibit much wider profiles with clear indication of line wings and possible outflows, mainly in the $^{12}$CO (4--3) and $^{12}$CO (3--2) transitions.
The $^{12}$CO (4--3) line is the strongest line for most sources, except for AR~6A and V960~Mon where both transitions are equally strong.
Indeed, for most FUors, the ratio between line profiles, ($J$=4--3)/($J$=3--2), is $<$1.5 at the systemic velocity of each object.
The two exceptions are FU~Ori, where the $J$=3--2 transition almost reaches 0 due to strong self-absorption, and V582~Aur, where there is a ratio of $\sim$4.5.
For the latter, this ratio suggests different excitation conditions, which can be explained by the intense radiation from two early B type stars within 30\,pc of V582~Aur that are exciting the region surrounding the FUor \citep{Kun_2017MNRAS4682325K}.

Some of our line profiles are different from those presented by \citet{Kospal_2017ApJ836226K}.
These discrepancies are because of differences in the distance to the FUors and in the shape of the apertures.
An example of the former is BBW~76, for which they used a distance 660\,pc larger than ours (see below), thus their aperture covered fewer pixels, causing a difference in the integrated flux of a factor of $\sim$3.
A similar scenario applies to AR~6A.
Concerning the different shapes of the apertures, \citet{Kospal_2017ApJ836226K} used a 10,000\,au circular aperture for all targets while we tailored the shape of our apertures.
Haro~5a~IRS, Reipurth~50 and V346~Nor are examples of this, where our apertures produced higher integrated fluxes by a factor of $\sim$3.

Unsurprisingly, we find that the $^{13}$CO~(3--2) transition produces the faintest line in all targets.
Its line profiles are single-peaked for most FUors with the maximum at velocities close to the systemic velocity (see below).
GM~Cha, iPTF~15afq, and OO~Ser are double-peaked with slightly less emission at the systemic velocity, a possible indication that $^{13}$CO~(3--2) is optically thick at the line center.

\section{Analysis}\label{sec:analysis}

\subsection{Distances}\label{ss:distances}
The estimation of gas masses is dependent on the distance to the target.
For FU~Ori, V899~Mon, AR~6A, V900~Mon, V960~Mon, and BBW~76 we used photogeometric distances from the Gaia Early Data Release 3 \citep{BailerJones2021AJ161147B}.
In the case of the more embedded objects (i.e.\ undetected by Gaia), L1551~IRS~5, V883~Ori, V1647~Ori, and V2775~Ori, we used the distances estimated from the distance to their molecular clouds and the positions of the FUors within them \citep[and references therein]{Connelley_2018ApJ861145C}.
For V582~Aur we used the distance estimated by \citet{Kun_2017MNRAS4682325K} under the assumption the FUor is related to the Aur~OB1 association.
We followed \citet{Tapia_2015MNRAS4464088T} and used the mean distance to the Great Carina Nebula (NGC~3372) for V723~Car.
The distance to iPTF~15afq was estimated by \citet{Park_2022ApJ941165P} after comparing different distance estimates based from kinematics, Gaia parallax and the distance to the CMa~OB1 association to which this object belongs.
For Reipurth~50, Z~CMa, GM~Cha, V346~Nor, and OO~Ser we used distances compiled from the literature \citep[and references therein]{Audard_2014prplconf387A}.
We compared our distances to those used by \citet{Kospal_2017ApJ836226K} and found that four FUors have different distance estimates: Haro~5a~IRS, AR~6A, V900~Mon and BBW~76.
The difference between our and their estimates are $-$79\,pc, 90\,pc, 30\,pc and $-$660\,pc, respectively.
If we had used the same apertures and velocity integration ranges as \citet{Kospal_2017ApJ836226K}, these differences in distance would translate to a difference in masses of factors of 0.69, 1.24, 1.06 and 0.37, respectively.

\subsection{Envelope masses}\label{ss:envelopes}
We used the $^{13}$CO~(3--2) emission to calculate the masses of the envelopes surrounding the FUors.
To calculate the integrated fluxes, we used the line profiles defined in \autoref{ss:line_profiles}, and integrated the channels that had emission above 3$\sigma$.
We assumed local thermodynamical equilibrium, used an excitation temperature of \SI{20}{\kelvin}, a \mbox{$^{13}$CO/$^{12}$CO} abundance ratio of 69 \citep{Wilson_1999RPPh62143W} and a \mbox{$^{12}$CO/H$_2$} abundance ratio of 10$^{-4}$ \citep{Bolatto_2013ARAA51207B}.
The velocity range used to calculate the line fluxes, the resulting line fluxes and the envelope mass estimates are presented in \autoref{tab:env_masses}.
To test the impact of our choice on gas temperature, we did the calculations using \SI{10}{\kelvin} or \SI{50}{\kelvin}, and found our estimated envelope masses would change by a factor of $\sim$0.94 or $\sim$2.55, respectively.
As we show later, some FUors have optically thick emission at velocities close to the systemic velocity, therefore, the estimated masses are lower limits for these sources.

\begin{deluxetable*}{lRRRRll}
\tablewidth{0pt}
  \tablecaption{Envelope masses of the FUors based on $^{13}$CO (3--2).  $v_{\min{}}$ and $v_{\max{}}$ indicate the velocity range used to integrate and calculate the line fluxes. The emission or absorption of the \SI{10}{\micro\meter} silicate feature is indicated, when known, and the reference used for each target.\label{tab:env_masses}}
\tablehead{\colhead{Name} & \colhead{$v_{\min{}}$} & \colhead{$v_{\max{}}$} & \colhead{Int. Flux} & \colhead{$M_{\mathrm{env}}$} & \colhead{\SI{10}{\micro\meter} feature} & \colhead{Si reference} \\
\colhead{} & \colhead{[\si{\kilo\meter\per\second}]} & \colhead{[\si{\kilo\meter\per\second}]} & \colhead{[\si{\jansky\kilo\meter\per\second}]} & \colhead{[M$_\odot$]} & \colhead{} & \colhead{}}
\startdata
L1551~IRS~5 &   3.11 &   9.03 & 4920.70 & 0.193 & Absorption & \citet{Quanz_2007ApJ668359Q}\\
V582~Aur    & -11.73 &  -8.34 &   43.35 & 0.137 & Emission  & \citet{Kospal_2020ApJ895L48K}\\
Haro~5a~IRS &   8.92 &  13.21 & 3714.50 & 1.032 & Absorption & \citet{Postel_2019AA631A30P}\\
V883~Ori    &   3.00 &   4.80 &  506.04 & 0.138 & Absorption & \citet{Quanz_2007ApJ668359Q}\\
Reipurth~50 &   1.28 &   7.61 & 1567.74 & 0.603 & Absorption & \citet{Quanz_2007ApJ668359Q}\\
FU~Ori      &  11.20 &  12.73 &  323.91 & 0.095 & Emission  & \citet{Quanz_2007ApJ668359Q}\\
V1647~Ori   &   7.95 &  11.79 & 1903.05 & 0.521 & Absorption & \citet{Quanz_2007ApJ668359Q}\\
V2775~Ori   &   0.72 &   4.53 & 1363.13 & 0.454 & Absorption & \citet{Kim_2016ApJS2268K}\\
V899~Mon    &   7.92 &  11.27 &  196.46 & 0.220 & Emission  & \citet{Kospal_2020ApJ895L48K}\\
AR~6a       &   0.24 &   8.40 &  899.39 & 1.295 & Unknown    & \\
V900~Mon    &  12.26 &  14.50 &   58.73 & 0.136 & Emission  & \citet{Kospal_2020ApJ895L48K}\\
V960~Mon    &  21.82 &  25.70 &   80.29 & 0.624 & Emission  & \citet{Kospal_2020ApJ895L48K}\\
Z~CMa       &  11.96 &  15.73 &  499.94 & 1.201 & Absorption & \citet{Quanz_2007ApJ668359Q}\\
iPTF~15afq  &  12.27 &  14.63 &   73.41 & 0.113 & Unknown    & \\
BBW~76      &  17.36 &  17.98 &    8.39 & 0.016 & Emission  & \citet{Quanz_2007ApJ668359Q}\\
V723~Car    & -21.70 & -16.13 &  294.20 & 3.341 & Absorption & \citet{Kospal_2020ApJ895L48K}\\
GM~Cha      &   3.31 &   6.39 & 4099.17 & 0.191 & Absorption & \citet{Manoj_2011ApJS19311M}\\
V346~Nor    &  -6.60 &  -0.86 &  443.84 & 0.395 & Absorption & \citet{Quanz_2007ApJ668359Q}\\
OO~Ser      &   5.22 &  12.38 & 3071.60 & 0.540 & Absorption & \citet{Quanz_2007ApJ668359Q}\\
HBC~687     &  16.71 &  17.40 &   34.74 & 0.010 & Emission  & \citet{Quanz_2007ApJ668359Q}\\
\enddata
\end{deluxetable*}

\subsection{Outflow detection}\label{ss:outflowdetection}
Here we explain the process we followed to determine whether a FUor had an outflow detection.
We inspected all the sources in our sample, including the ones that \citet{Kospal_2017ApJ836226K} considered as \emph{not} having an outflow.

As mentioned earlier, high-velocity wings in the line profiles of $^{12}$CO are a common indicator of outflows, and these are present in some of our FUors.
However, this feature by itself is not enough.
Thus, we examined the $^{12}$CO channel maps for each target (found in \autoref{app:channelmaps}) to verify the existence of the outflows via a visual inspection of the different distribution of the gas between the channels close to the systemic velocity and outwards to higher velocities.
The envelope emission dominates the velocities closest to the systemic, which are approximately the same velocities covered by the $^{13}$CO emission (\autoref{tab:env_masses}), so we focused on velocities beyond these.
If there emission is not detected at velocities higher the ones overran by the envelope then we consider the FUor as not having an outflow.
As outflows originate from the protostars, it is expected that at lower velocities (with respect to the systemic) the outflow is extended and its position is closer to the FUor.
So in the case there is emission in the maps beyond the envelope velocities, we checked the separation of this gas with respect the position of the FUor.
Should the emission be close to the FUor we consider them to be an outflow, and in the case they are separated we do not.
When compared to the outflow detections of \citet{Kospal_2017ApJ836226K}, our methodology resulted in almost the same detections and non-detections with the only difference begin V900~Mon.
In this case, the line profiles do not show high velocity wings and thus resulted in a non-detection for them.

After we detected an outflow, we determined the velocity ranges at which it was present in the blueshifted and redshifted sides.
First, we found the velocity channel on which the envelope is not dominant and considered it as the ``inner'' velocity, $v_\mathrm{in}$, of the outflow.
Next, we located the velocity channel where a 3$\sigma$ detection was not found and considered it as the ``outer'' velocity, $v_\mathrm{out}$.
We then calculated the maximum velocity of each lobe of the outflow with respect to the systemic velocity as $v_{\max{}} = v_\mathrm{out}-v_{sys}$.
The list of FUors with outflows and these three velocities are presented in \autoref{tab:outflows_geometry}.
In \autoref{fig:spectra} we marked with vertical dotted lines the velocity ranges where outflows are detected.
For two FUors, V900~Mon and iPTF~15afq, the $^{12}$CO (4--3) emission at velocities close to the systemic does not appear to be dominated by the envelope, thus, we used the systemic as $v_\mathrm{in}$ for both lobes.

Finally, we used these velocities to produce blueshifted and redshifted integrated emission maps of the J=3--2 and J=4--3 transitions of $^{12}$CO, and their contour maps are presented in \autoref{fig:outflows} and \autoref{fig:outflows43}.
We used these maps to estimate the position angle of the outflow, reported in \autoref{tab:outflows_geometry}, and to estimate the extension of each lobe (see below).

\begin{deluxetable*}{lCClrrcrrrrcrr}
\tablecaption{Position angles, velocities, extensions and dynamical times of outflows\label{tab:outflows_geometry}}
\tablehead{
  & & & & \multicolumn{5}{c}{CO (3--2)} & \multicolumn{5}{c}{CO (4--3)}\\
  \cmidrule(l{2pt}r{2pt}){5-9}  \cmidrule(l{2pt}r{2pt}){10-14}
  \colhead{Target}  & \colhead{Inc.} & \colhead{PA} & \colhead{Side} & \colhead{$v_\mathrm{in}$} & \colhead{$v_\mathrm{out}$} & \colhead{$|v_{\max{}}|$} & \colhead{$R_\mathrm{lobe}$} & \colhead{$\tau_\mathrm{d}$} & \colhead{$v_\mathrm{in}$} & \colhead{$v_\mathrm{out}$} & \colhead{$|v_{\max{}}|$} & \colhead{$R_\mathrm{lobe}$} & \colhead{$\tau_\mathrm{d}$}\\
  &  \colhead{[$^\circ$]} &\colhead{[$^\circ$]} & & \colhead{[\si[per-mode=fraction]{\kilo\meter\per\second}]} & \colhead{[\si[per-mode=fraction]{\kilo\meter\per\second}]}& \colhead{[\si[per-mode=fraction]{\kilo\meter\per\second}]} & \colhead{[10$^3$\,\si{\astronomicalunit}]}  & \colhead{[10$^{3}$\,\si{\year}]} & \colhead{[\si[per-mode=fraction]{\kilo\meter\per\second}]} & \colhead{[\si[per-mode=fraction]{\kilo\meter\per\second}]}& \colhead{[\si[per-mode=fraction]{\kilo\meter\per\second}]} & \colhead{[10$^3$\,\si{\astronomicalunit}]}  & \colhead{[10$^{3}$\,\si{\year}]}
}
\startdata
L1551~IRS~5 & 70 & 65  & Blue* & \num{5.03}  & \num{-0.27}  & \num{6.73} & \num{11.7} & \num{8.2}   & \num{5.64}  & \num{0.93}  & \num{5.54} & \num{11.7} & \num{10.0}\\
            &    &     & Red*  & \num{7.57}  & \num{12.80}  & \num{6.33} & \num{11.7} & \num{8.7}   & \num{7.67}  & \num{12.24} & \num{5.77} & \num{11.7} & \num{9.6}\\
Haro~5a~IRS & 50 & 70  & Blue* & \num{9.20}  & \num{5.50}   & \num{5.35} & \num{27.2} & \num{24.1}  & \num{9.53}  & \num{5.30}  & \num{5.55} & \num{25.3} & \num{21.6}\\
            &    &     & Red*  & \num{13.40} & \num{16.63}  & \num{5.78} & \num{26.6} & \num{21.8}  & \num{12.80} & \num{15.50} & \num{4.65} & \num{22.9} & \num{23.3}\\
Reipurth~50 & 70 & 150 & Blue* & \num{1.03}  & \num{-0.98}  & \num{4.81} & \num{18.8} & \num{18.5}  & \num{2.60}  & \num{1.01}  & \num{2.82} & \num{9.2}  & \num{15.5}\\
            &    &     & Red*  & \num{7.91}  & \num{8.97}   & \num{5.14} & \num{32.0} & \num{29.5}  & \num{5.18}  & \num{8.01}  & \num{4.18} & \num{23.6} & \num{26.7}\\
V2775~Ori   & 10 & -   & Blue  & \num{1.36}  & \num{-5.58}  & \num{8.66} & \num{6.3}  & \num{3.4}   & \num{1.61}  & \num{-1.62} & \num{4.70} & \num{10.7} & \num{10.8}\\
            &    &     & Red   & \num{4.57}  & \num{8.24}   & \num{5.16} & \num{4.8}  & \num{4.4}   & \num{4.34}  & \num{7.01}  & \num{3.93} & \num{7.1}  & \num{8.5}\\
V899~Mon    & 50 & 60  & Blue* & \num{8.60}  & \num{6.92}   & \num{2.62} & \num{46.1} & \num{83.3}  & \num{8.93}  & \num{7.99}  & \num{1.55} & \num{24.4} & \num{74.6}\\
            &    &     & Red*  & \num{10.85} & \num{12.47}  & \num{2.93} & \num{23.0} & \num{37.2}  & \num{10.71} & \num{11.66} & \num{2.12} & \num{18.4} & \num{41.2}\\
V900~Mon$^\dag{}$ & 30 & 80  & Blue  & \num{12.02} & \num{11.22}  & \num{2.53} & \num{21.5} & \num{40.4}  & \num{13.44} & \num{12.49} & \num{1.26} & \num{21.5} & \num{80.9}\\
            &    &     & Red   & \num{14.46} & \num{14.73}  & \num{0.98} & \num{25.9} & \num{125.7} & \num{13.44} & \num{14.23} & \num{0.48} & \num{25.9} & \num{256.2}\\
V960~Mon    & 10 & -   & Blue  & \num{21.42} & \num{17.32}  & \num{6.48} & \num{14.5} & \num{10.6}  & \num{22.12} & \num{20.93} & \num{2.87} & \num{14.5} & \num{23.9}\\
            &    &     & Red   & \num{27.14} & \num{28.80}  & \num{5.00} & \num{14.5} & \num{13.7}  & \num{24.61} & \num{27.04} & \num{3.24} & \num{14.5} & \num{21.2}\\
Z~CMa       & 30 & 45  & Blue  & \num{12.13} & \num{10.71}  & \num{3.15} & \num{29.9} & \num{45.0}  & \num{12.44} & \num{11.25} & \num{2.61} & \num{21.6} & \num{39.2}\\
            &    &     & Red   & \num{16.00} & \num{19.34}  & \num{5.48} & \num{47.0} & \num{40.6}  & \num{15.56} & \num{19.18} & \num{5.32} & \num{40.8} & \num{36.3}\\
iPTF~15afq$^\dag{}$  & 50 & 135 & Blue* & \num{11.16} & \num{9.84}   & \num{4.20} & \num{28.1} & \num{31.7}  & \num{13.00} & \num{11.36} & \num{2.89} & \num{28.1} & \num{49.7}\\
            &    &     & Red*  & \num{17.97} & \num{20.65}  & \num{6.61} & \num{60.4} & \num{43.3}  & \num{13.00} & \num{16.22} & \num{1.97} & \num{60.4} & \num{131.3}\\
GM~Cha      & 70 & 100 & Blue  & \num{3.48}  & \num{2.82}   & \num{2.02} & \num{2.8}  & \num{6.5}   & \num{3.65}  & \num{3.00}  & \num{1.84} & \num{1.3}  & \num{3.3}\\
            &    &     & Red   & \num{6.59}  & \num{10.00}  & \num{5.16} & \num{6.8}  & \num{6.2}   & \num{6.08}  & \num{9.80}  & \num{4.96} & \num{6.4}  & \num{6.1}\\
V346~Nor    & 30 & 45  & Blue  & \num{-5.41} & \num{-11.16} & \num{8.18} & \num{22.7} & \num{13.1}  & \num{-4.58} & \num{-9.59} & \num{6.61} & \num{18.0} & \num{12.9}\\
            &    &     & Red*  & \num{-0.25} & \num{5.94}   & \num{8.92} & \num{39.6} & \num{21.1}  & \num{-1.26} & \num{4.38}  & \num{7.36} & \num{30.0} & \num{19.3}
\enddata
  \tablecomments{The inclination angles here are the values used for the inclination correction in \autoref{ss:outflow_parameters}, see text for details. The position angles were estimated by hand using the CO (3--2) integrated emission maps. The asterisk (*) indicates the lobes that extend beyond the field-of-view of our observations, thus their values of $R_\mathrm{lobe}$ and $\tau_d$ are lower limits. The $\dag{}$ labels the two FUors with tentative outflow detections.}
\end{deluxetable*}

\subsection{Outflow properties}\label{ss:outflowproperties}
One of the goals of this work is to compare the outflows emanating from eruptive young stars to those from quiescent young stellar objects.
We carried out our calculations for the blueshifted and redshifted parts of the spectra separately and here we describe how we carried out these calculations.
The results for the J=3--2 and J=4--3 transitions are presented in \autoref{tab:outflows_32} and \autoref{tab:outflows_43}, respectively.

\subsubsection{Mass, momenta, and energy}
The outflow masses were calculated assuming the wind emission is in local thermodynamical equilibrium with an excitation temperature of \SI{75}{\kelvin} \citep{vanKempen_2009AA508259V,Yildiz_2015AA576A109Y} and assuming a CO abundance of 10$^{-4}$ with respect to H$_2$ \citep{Bolatto_2013ARAA51207B}.
We calculated the mass ($M_v$) for each velocity channel ($v$) for all pixels above 3$\sigma$.
Afterwards, we calculated the momentum and kinematic energy for each channel with $P_v = M_v \times v$ and $E_v = 0.5 M_v \times v^2$, respectively.
Finally, we integrated the three properties over the same velocity range to obtain the total values ($M_\mathrm{of}$, $P_\mathrm{of}$, $E_\mathrm{of}$) for the blueshifted and redshifted lobes of the outflow.

\subsubsection{Force and luminosity}
The outflow force and luminosity are calculated as $F_\mathrm{of} = P_\mathrm{of} / \tau_d$ and $L_\mathrm{of} = E_\mathrm{of} / \tau_d$, respectively, where $\tau_d$ is the dynamical time of the outflow.
The dynamical time is defined as $\tau_d = R_\mathrm{lobe} / v_{\max{}}$, where $R_\mathrm{lobe}$ is the projected extension of the outflow lobe and $v_{\max{}}$ is the maximum velocity of the outflow.

For the projected extension of the outflows, we used the integrated emission maps (\autoref{fig:outflows} and \autoref{fig:outflows43}) to measure the separation between the position of the star and the maximum length at which the outflow is above 3$\upsigma$ for each transition separately.
However, the outflows around some of our targets extend beyond the field of view of our observations so our extension measurements are only a lower limit.
The lobes for which this is the case are indicated with an asterisk in \autoref{tab:outflows_geometry}.
We estimated the maximum outflow velocity for each lobe independently by calculating the difference between the systemic velocity and the minimum/maximum velocity where there is blueshifted/redshifted emission.
In both $R_\mathrm{lobe}$ and $v_{\max{}}$, the sensitivity and spatial resolution of the observations directly affect their measured values.

\begin{deluxetable*}{lccccCC}  
\tablecaption{Outflow properties from $^{12}$CO~(3--2) observations assuming they are optically thin.\label{tab:outflows_32}}
\tablehead{
\colhead{Target}  & \colhead{Side} & \colhead{$M_\mathrm{of}$} & \colhead{$P_\mathrm{of}$} & \colhead{$E_\mathrm{of}$} & \colhead{$F_\mathrm{of}$}  & \colhead{$L_\mathrm{of}$}\\
  & & \colhead{[M$_\odot$]} & \colhead{[M$_\odot$ \si{\kilo\meter\per\second}]} & \colhead{[erg]} & \colhead{[M$_\odot$ yr$^{-1}$ \si{\kilo\meter\per\second}]} & \colhead{[L$_\odot$]}
}
\startdata
L1551~IRS~5         & Blue & \num{6.3e-03} & \num{1.9e-02} & \num{6.8e+41} & \num{2.4e-06} & \num{6.9e-04}\\
                    & Red  & \num{6.8e-03} & \num{1.7e-02} & \num{4.9e+41} & \num{1.9e-06} & \num{4.7e-04}\\
Haro~5a~IRS         & Blue & \num{1.6e-02} & \num{3.6e-02} & \num{8.7e+41} & \num{1.5e-06} & \num{3.0e-04}\\
                    & Red  & \num{8.6e-03} & \num{2.7e-02} & \num{8.6e+41} & \num{1.2e-06} & \num{3.2e-04}\\
Reipurth~50         & Blue & \num{3.5e-03} & \num{1.1e-02} & \num{3.6e+41} & \num{6.0e-07} & \num{1.6e-04}\\
                    & Red  & \num{7.2e-03} & \num{3.3e-02} & \num{1.5e+42} & \num{1.1e-06} & \num{4.4e-04}\\
V2775~Ori           & Blue & \num{4.6e-02} & \num{1.7e-01} & \num{7.4e+42} & \num{4.9e-05} & \num{1.8e-02}\\
                    & Red  & \num{3.5e-02} & \num{9.5e-02} & \num{2.8e+42} & \num{2.2e-05} & \num{5.3e-03}\\
V899~Mon            & Blue & \num{1.3e-02} & \num{2.0e-02} & \num{3.2e+41} & \num{2.4e-07} & \num{3.2e-05}\\
                    & Red  & \num{2.2e-02} & \num{3.8e-02} & \num{6.8e+41} & \num{1.0e-06} & \num{1.5e-04}\\
V900~Mon$^\dag{}$   & Blue & \num{6.0e-03} & \num{1.2e-02} & \num{2.5e+41} & \num{3.1e-07} & \num{5.3e-05}\\
                    & Red  & \num{1.3e-02} & \num{1.0e-02} & \num{8.0e+40} & \num{7.9e-08} & \num{5.2e-06}\\
V960~Mon            & Blue & \num{1.1e-01} & \num{4.2e-01} & \num{1.7e+43} & \num{3.9e-05} & \num{1.4e-02}\\
                    & Red  & \num{5.9e-02} & \num{2.3e-01} & \num{8.7e+42} & \num{1.6e-05} & \num{5.2e-03}\\
Z~CMa               & Blue & \num{1.9e-02} & \num{4.2e-02} & \num{9.3e+41} & \num{9.4e-07} & \num{1.7e-04}\\
                    & Red  & \num{4.7e-02} & \num{1.5e-01} & \num{4.8e+42} & \num{3.5e-06} & \num{9.7e-04}\\
iPTF~15afq$^\dag{}$ & Blue & \num{1.8e-02} & \num{5.8e-02} & \num{1.9e+42} & \num{1.8e-06} & \num{5.0e-04}\\
                    & Red  & \num{3.8e-02} & \num{1.8e-01} & \num{8.9e+42} & \num{4.2e-06} & \num{1.7e-03}\\
GM~Cha              & Blue & \num{7.4e-04} & \num{1.2e-03} & \num{1.8e+40} & \num{1.8e-07} & \num{2.3e-05}\\
                    & Red  & \num{1.2e-03} & \num{3.3e-03} & \num{9.8e+40} & \num{5.3e-07} & \num{1.3e-04}\\
V346~Nor            & Blue & \num{1.7e-02} & \num{6.4e-02} & \num{2.7e+42} & \num{4.8e-06} & \num{1.7e-03}\\
                    & Red  & \num{4.6e-02} & \num{2.0e-01} & \num{9.8e+42} & \num{9.8e-06} & \num{3.9e-03}
\enddata
\tablecomments{The $\dag{}$ labels the two FUors with tentative outflow detections.}
\end{deluxetable*}

\begin{deluxetable*}{lccccCC}  
\tablecaption{Outflow properties estimated from the $^{12}$CO~(4--3) observations.\label{tab:outflows_43}}
\tablehead{
\colhead{Target}  & \colhead{Side} & \colhead{$M_\mathrm{of}$} & \colhead{$P_\mathrm{of}$} & \colhead{$E_\mathrm{of}$} & \colhead{$F_\mathrm{of}$}  & \colhead{$L_\mathrm{of}$}\\
  & & \colhead{[M$_\odot$]} & \colhead{[M$_\odot$ \si{\kilo\meter\per\second}]} & \colhead{[erg]} & \colhead{[M$_\odot$ yr$^{-1}$ \si{\kilo\meter\per\second}]} & \colhead{[L$_\odot$]}
}
\startdata
L1551~IRS~5         & Blue & \num{3.8e-03} & \num{8.4e-03} & \num{2.3e+41} & \num{8.4e-07} & \num{1.9e-04}\\
                    & Red  & \num{2.9e-03} & \num{6.9e-03} & \num{1.9e+41} & \num{7.2e-07} & \num{1.6e-04}\\
Haro~5a~IRS         & Blue & \num{1.8e-02} & \num{3.3e-02} & \num{6.7e+41} & \num{1.6e-06} & \num{2.6e-04}\\
                    & Red  & \num{1.3e-02} & \num{2.9e-02} & \num{6.9e+41} & \num{1.2e-06} & \num{2.4e-04}\\
Reipurth~50         & Blue & \num{4.3e-03} & \num{7.3e-03} & \num{1.3e+41} & \num{4.6e-07} & \num{6.8e-05}\\
                    & Red  & \num{1.9e-02} & \num{4.5e-02} & \num{1.2e+42} & \num{1.7e-06} & \num{3.7e-04}\\
V2775~Ori           & Blue & \num{9.9e-03} & \num{2.4e-02} & \num{6.4e+41} & \num{2.2e-06} & \num{4.9e-04}\\
                    & Red  & \num{1.0e-02} & \num{2.3e-02} & \num{5.7e+41} & \num{2.7e-06} & \num{5.5e-04}\\
V899~Mon            & Blue & \num{3.2e-03} & \num{3.3e-03} & \num{3.7e+40} & \num{4.5e-08} & \num{4.2e-06}\\
                    & Red  & \num{2.9e-03} & \num{4.5e-03} & \num{7.3e+40} & \num{1.1e-07} & \num{1.4e-05}\\
V900~Mon$^\dag{}$   & Blue & \num{1.8e-02} & \num{1.3e-02} & \num{1.1e+41} & \num{1.7e-07} & \num{1.2e-05}\\
                    & Red  & \num{1.3e-02} & \num{2.8e-03} & \num{8.1e+39} & \num{1.1e-08} & \num{2.5e-07}\\
V960~Mon            & Blue & \num{4.0e-02} & \num{8.2e-02} & \num{1.7e+42} & \num{3.4e-06} & \num{6.0e-04}\\
                    & Red  & \num{1.7e-01} & \num{2.6e-01} & \num{4.5e+42} & \num{1.2e-05} & \num{1.7e-03}\\
Z~CMa               & Blue & \num{1.6e-02} & \num{2.8e-02} & \num{4.9e+41} & \num{7.2e-07} & \num{1.1e-04}\\
                    & Red  & \num{4.3e-02} & \num{1.1e-01} & \num{3.2e+42} & \num{3.0e-06} & \num{7.2e-04}\\
iPTF~15afq$^\dag{}$ & Blue & \num{2.2e-02} & \num{3.6e-02} & \num{6.3e+41} & \num{7.2e-07} & \num{1.1e-04}\\
                    & Red  & \num{2.6e-02} & \num{2.2e-02} & \num{2.7e+41} & \num{1.7e-07} & \num{1.7e-05}\\
GM~Cha              & Blue & \num{3.8e-04} & \num{5.3e-04} & \num{7.6e+39} & \num{1.6e-07} & \num{1.9e-05}\\
                    & Red  & \num{9.1e-04} & \num{2.1e-03} & \num{5.6e+40} & \num{3.4e-07} & \num{7.5e-05}\\
V346~Nor            & Blue & \num{1.2e-02} & \num{3.7e-02} & \num{1.3e+42} & \num{2.8e-06} & \num{8.4e-04}\\
                    & Red  & \num{2.5e-02} & \num{8.5e-02} & \num{3.3e+42} & \num{4.5e-06} & \num{1.4e-03}
\enddata
\tablecomments{The $\dag{}$ labels the two FUors with tentative outflow detections.}
\end{deluxetable*}

\subsubsection{Caveats}\label{sss:caveats}
Due to the nature of our observations and methodology, it is important to understand the limitations of our estimated values.

First, the large uncertainties in the estimations of the outflow masses due to the presence of envelopes.
This surrounding material dominates emission at velocities close to the systemic velocity, thus we have calculated the outflow properties using only channels where the outflow is the predominant flux contributor.
Therefore, we are knowingly underestimating the outflow masses by not integrating the emission at low-velocities to prevent this contamination.
However, it is likely that some of the envelope emission is still included into our calculations.
Indeed, in a couple of cases (V883~Ori and V1647~Ori) we are not able to separate their known outflows due to their emission being at velocities comparable to those of the surrounding cloud.

Second, the sensitivity of the observations put strong constraints on the maximum velocities where outflows are detected.
For example, in the case of L1551~IRS~5, \citet{Yildiz_2015AA576A109Y} reported a sum of maximum velocities of \kms{21.5}, while we obtained $\sim$\kms{13.1} (see \autoref{tab:outflows_geometry}).
Their JCMT observations had typical noise values $\lesssim$\SI{0.1}{\kelvin}, while our APEX observations have a noise value of $\sim$\SI{0.38}{\kelvin}.

\begin{figure*}  
\centering
\includegraphics[width=\textwidth]{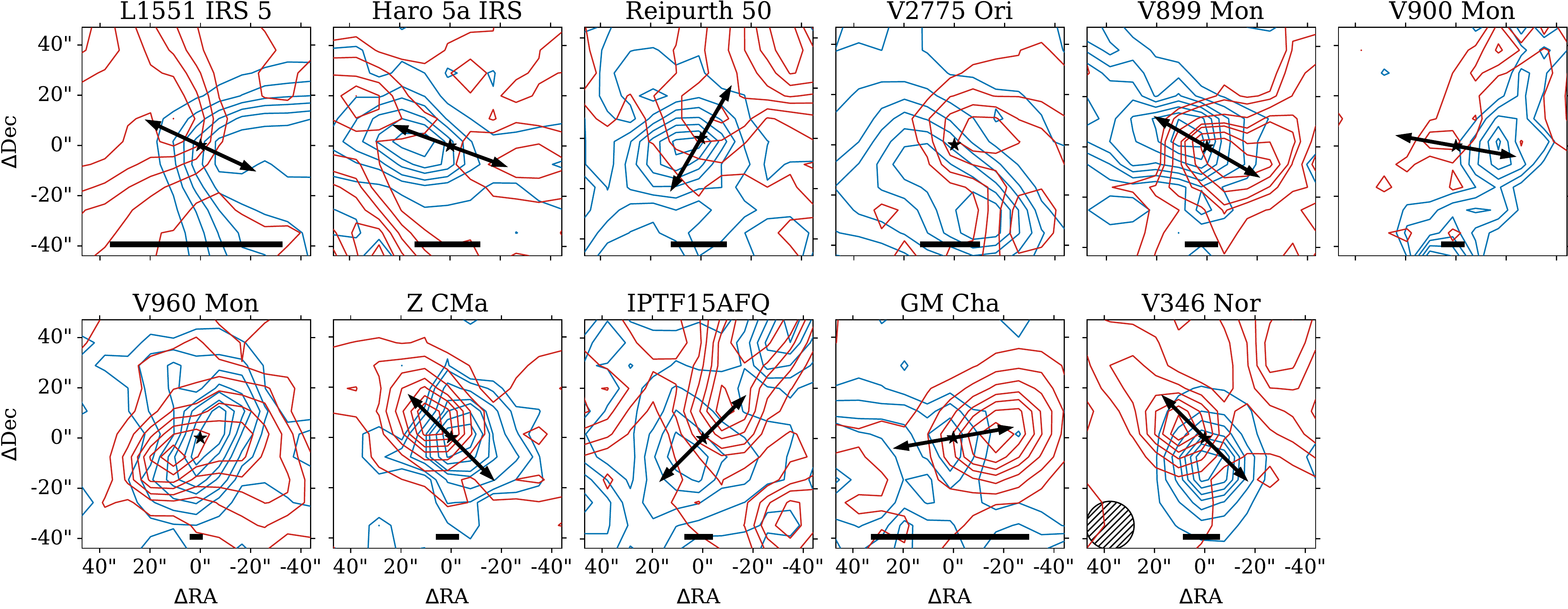}
\caption{The red and blue contours show redshifted and blueshifted CO (J=3--2) emission integrated in the velocity ranges indicated in \autoref{fig:spectra}.
The star symbols mark the stellar position as given in \autoref{tab:values}.
The hatched circle in the bottom right frame is the APEX beam size and the arrows indicate the orientation of the outflow.
In the cases of V2775~Ori and V960~Mon, the outflow appears to be expanding in the direction of the line of sight.
For L1551~IRS~5, the blueshifted contours are 3, 13, 24, 35, 45, 56, 67 and 78$\upsigma$ for $\upsigma$=\SI{0.38}{\kelvin\kilo\meter\per\second} while the redshifted contours are 3, 18, 33, 48, 63, 78, 93 and 109$\upsigma$ for $\upsigma$=\SI{0.33}{\kelvin\kilo\meter\per\second}.
For Haro~5a~IRS, the blueshifted contours are 3, 7, 12, 16, 21, 25, 30 and 35$\upsigma$ for $\upsigma$=\SI{0.61}{\kelvin\kilo\meter\per\second} while the redshifted contours are 3, 6, 10, 14, 18, 22, 26 and 30$\upsigma$ for $\upsigma$=\SI{0.48}{\kelvin\kilo\meter\per\second}.
For Reipurth~50, the blueshifted contours are 3, 5, 7, 9, 11, 13, 15 and 17$\upsigma$ for $\upsigma$=\SI{0.48}{\kelvin\kilo\meter\per\second} while the redshifted contours are 3, 6, 10, 14, 17, 21, 25 and 29$\upsigma$ for $\upsigma$=\SI{0.43}{\kelvin\kilo\meter\per\second}.
For V2775~Ori, the blueshifted contours are 3, 12, 22, 31, 41, 50, 60 and 70$\upsigma$ for $\upsigma$=\SI{0.73}{\kelvin\kilo\meter\per\second} while the redshifted contours are 3, 7, 12, 17, 21, 26, 31 and 36$\upsigma$ for $\upsigma$=\SI{0.68}{\kelvin\kilo\meter\per\second}.
For V899~Mon, the blueshifted contours are 3, 5, 7, 9, 11, 13, 15 and 17$\upsigma$ for $\upsigma$=\SI{0.52}{\kelvin\kilo\meter\per\second} while the redshifted contours are 3, 4, 6, 8, 9, 11, 13 and 15$\upsigma$ for $\upsigma$=\SI{0.43}{\kelvin\kilo\meter\per\second}.
For V900~Mon, the blueshifted contours are 3, 4, 5, 6, 7, 8, 9 and 10$\upsigma$ for $\upsigma$=\SI{0.62}{\kelvin\kilo\meter\per\second} while the redshifted contours are 3, 4, 5, 6, 7 and 8$\upsigma$ for $\upsigma$=\SI{0.47}{\kelvin\kilo\meter\per\second}.
For V960~Mon, the blueshifted contours are 3, 6, 9, 13, 16, 20, 23 and 27$\upsigma$ for $\upsigma$=\SI{0.44}{\kelvin\kilo\meter\per\second} while the redshifted contours are 3, 7, 11, 15, 19, 23, 27 and 32$\upsigma$ for $\upsigma$=\SI{0.25}{\kelvin\kilo\meter\per\second}.
For Z~CMa, the blueshifted contours are 3, 4, 6, 8, 9, 11, 13 and 15$\upsigma$ for $\upsigma$=\SI{0.52}{\kelvin\kilo\meter\per\second} while the redshifted contours are 3, 8, 13, 19, 24, 30, 35 and 41$\upsigma$ for $\upsigma$=\SI{0.39}{\kelvin\kilo\meter\per\second}.
For iPTF~15afq, the blueshifted contours are 3, 5, 7, 9, 11, 13, 15 and 17$\upsigma$ for $\upsigma$=\SI{0.44}{\kelvin\kilo\meter\per\second} while the redshifted contours are 3, 6, 10, 13, 17, 20, 24 and 28$\upsigma$ for $\upsigma$=\SI{0.32}{\kelvin\kilo\meter\per\second}.
For GM~Cha, the blueshifted contours are 3, 5, 7, 9, 11, 13, 15 and 18$\upsigma$ for $\upsigma$=\SI{0.43}{\kelvin\kilo\meter\per\second} while the redshifted contours are 3, 9, 16, 23, 29, 36, 43 and 50$\upsigma$ for $\upsigma$=\SI{0.39}{\kelvin\kilo\meter\per\second}.
For V346~Nor, the blueshifted contours are 3, 12, 22, 32, 41, 51, 61 and 71$\upsigma$ for $\upsigma$=\SI{0.52}{\kelvin\kilo\meter\per\second} while the redshifted contours are 3, 11, 20, 29, 38, 47, 56 and 65$\upsigma$ for $\upsigma$=\SI{0.56}{\kelvin\kilo\meter\per\second}.%
\label{fig:outflows}}
\end{figure*}

\begin{figure*}[!ht]  
\centering
\includegraphics[width=\textwidth]{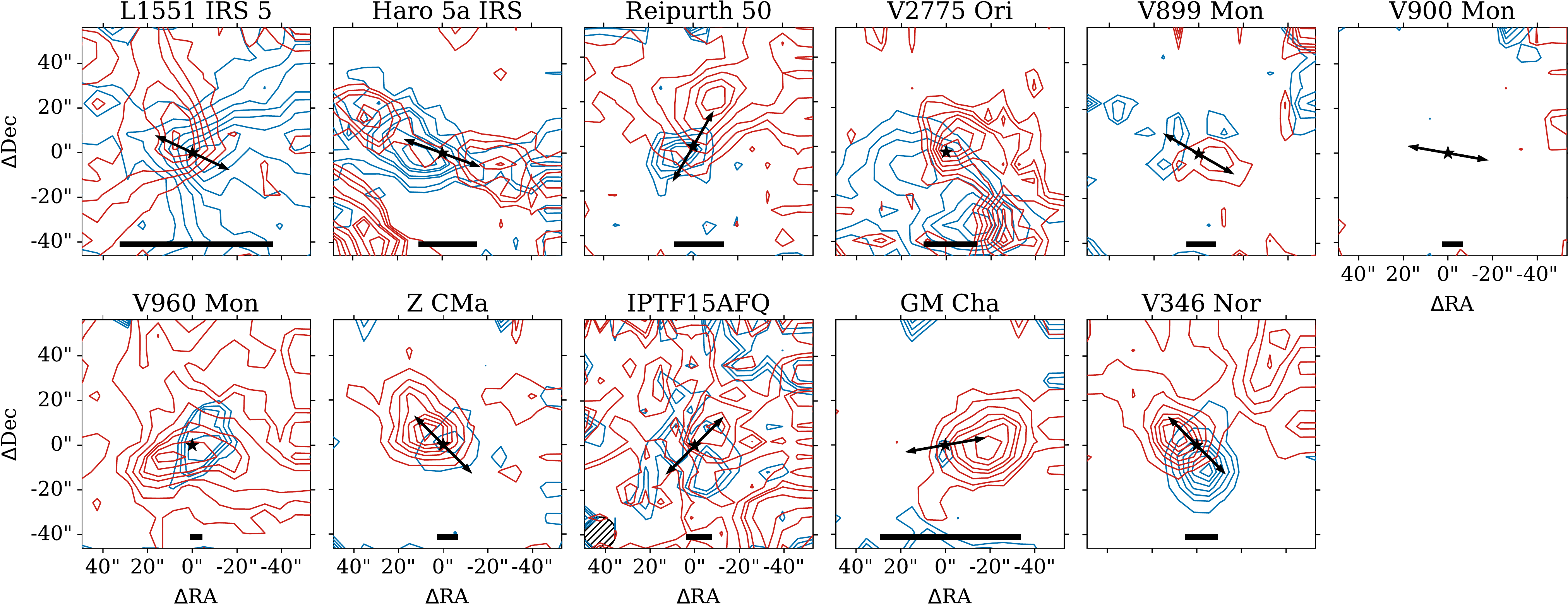}
  \caption{Similar to \autoref{fig:outflows} but for $^{12}$CO~(4--3).
  In the case of V900~Mon, the emission at this transition is not significant enough to be seen in this map and the emission detected is in the outer parts of the field of view. In the iPTF~15afq map, the emission of the outflow is weak compared to the surrounding gas.
For L1551~IRS~5, the blueshifted contours are 4, 7, 11, 15, 19, 23, 27 and 31$\upsigma$ for $\upsigma$=\SI{0.96}{\kelvin\kilo\meter\per\second} while the redshifted contours are 4, 8, 12, 17, 21, 26, 30 and 35$\upsigma$ for $\upsigma$=\SI{0.84}{\kelvin\kilo\meter\per\second}.
For Haro~5a~IRS, the blueshifted contours are 4, 5, 7, 8, 10, 11, 13 and 15$\upsigma$ for $\upsigma$=\SI{2.09}{\kelvin\kilo\meter\per\second} while the redshifted contours are 4, 5, 6, 8, 9, 11, 12 and 14$\upsigma$ for $\upsigma$=\SI{1.65}{\kelvin\kilo\meter\per\second}.
For Reipurth~50, the blueshifted contours are 5, 6, 7, 8 and 9$\upsigma$ for $\upsigma$=\SI{1.22}{\kelvin\kilo\meter\per\second} while the redshifted contours are 5, 7, 10, 13, 16, 19, 22 and 25$\upsigma$ for $\upsigma$=\SI{1.13}{\kelvin\kilo\meter\per\second}.
For V2775~Ori, the blueshifted contours are 5, 8, 11, 15, 18, 22, 25 and 29$\upsigma$ for $\upsigma$=\SI{1.72}{\kelvin\kilo\meter\per\second} while the redshifted contours are 5, 6, 7, 8, 9, 10, 11 and 12$\upsigma$ for $\upsigma$=\SI{1.62}{\kelvin\kilo\meter\per\second}.
For V899~Mon, the blueshifted contours are 3, 4, 5, 6, 7, 8 and 9$\upsigma$ for $\upsigma$=\SI{1.50}{\kelvin\kilo\meter\per\second} while the redshifted contours are 3, 4, 5, 6, 7 and 8$\upsigma$ for $\upsigma$=\SI{1.22}{\kelvin\kilo\meter\per\second}.
For V900~Mon, the blueshifted contours are 4, 5, 6, 7 and 8$\upsigma$ for $\upsigma$=\SI{2.16}{\kelvin\kilo\meter\per\second} while the redshifted contours are 4, 5 and 6$\upsigma$ for $\upsigma$=\SI{1.64}{\kelvin\kilo\meter\per\second}.
For V960~Mon, the blueshifted contours are 4, 5, 6 and 7$\upsigma$ for $\upsigma$=\SI{1.38}{\kelvin\kilo\meter\per\second} while the redshifted contours are 4, 6, 9, 12, 14, 17, 20 and 23$\upsigma$ for $\upsigma$=\SI{0.82}{\kelvin\kilo\meter\per\second}.
For Z~CMa, the blueshifted contours are 4, 7, 10, 13, 16, 19, 22 and 26$\upsigma$ for $\upsigma$=\SI{1.26}{\kelvin\kilo\meter\per\second} while the redshifted contours are 4, 7, 10, 13, 16, 19, 22 and 25$\upsigma$ for $\upsigma$=\SI{0.94}{\kelvin\kilo\meter\per\second}.
For iPTF~15afq, the blueshifted contours are 4, 5, 6, 8, 9, 11, 12 and 14$\upsigma$ for $\upsigma$=\SI{1.15}{\kelvin\kilo\meter\per\second} while the redshifted contours are 4, 5, 7, 9, 10, 12, 14 and 16$\upsigma$ for $\upsigma$=\SI{0.88}{\kelvin\kilo\meter\per\second}.
For GM~Cha, the blueshifted contours are 3, 4, 5, 6 and 7$\upsigma$ for $\upsigma$=\SI{1.21}{\kelvin\kilo\meter\per\second} while the redshifted contours are 4, 6, 9, 12, 14, 17, 20 and 23$\upsigma$ for $\upsigma$=\SI{1.09}{\kelvin\kilo\meter\per\second}.
For V346~Nor, the blueshifted contours are 4, 8, 13, 18, 23, 28, 33 and 38$\upsigma$ for $\upsigma$=\SI{1.66}{\kelvin\kilo\meter\per\second} while the redshifted contours are 4, 8, 12, 16, 20, 24, 28 and 33$\upsigma$ for $\upsigma$=\SI{1.77}{\kelvin\kilo\meter\per\second}.%
  \label{fig:outflows43}}
\end{figure*}

Finally, the extension of the outflow can reach beyond the field of view of our observations.
An extreme example is that L1551~IRS~5 whose outflow extends out to $\sim$20$'$ \citep[e.g.][]{Stojimirovic_2006ApJ649280S} and our field of view is less than 1$'$ (see \autoref{fig:outflows} and \autoref{fig:outflows43}).

Therefore, both the outflow masses and their dynamical ages should be considered as lower limits, while the outflow forces and luminosities must be considered as highly uncertain.

\subsection{Optical depth correction}\label{ss:outflows}
We calculated the outflow parameters assuming the $^{12}$CO lines are optically thin, however, this isotopologue is typically optically thick.
One way to correct for this optical depth issue is by using the $^{13}$CO emission, under the assumption that \emph{that} isotopologue is optically thin, and use it to correct the fluxes of $^{12}$CO\@.
In this section we present our methodology to correct the emission of the J=3--2 transition of the $^{12}$CO\@.

This correction was done following the procedure presented by \citet{Dunham_2014ApJ78329D}.
We assumed that both CO isotopologues are in local thermodynamical equilibrium at the same excitation temperature, and with identical beam filling factors.
Under these conditions, the brightness temperature ratio between the two isotopologues is given by
\begin{equation}\label{eq:tmbratio}
\frac{T_\mathrm{mb,12}}{T_\mathrm{mb,13}} = \frac{1-e^{-\tau_\mathrm{12}}}{1-e^{-\tau_\mathrm{13}}},
\end{equation}
where $T_\mathrm{mb,12}$ and $T_\mathrm{mb,13}$ are the brightness temperatures of $^{12}$CO and $^{13}$CO, respectively, and $\tau_\mathrm{12}$ and $\tau_\mathrm{13}$ are their respective opacities.
Assuming that $^{13}$CO is optically thin, \autoref{eq:tmbratio} can be re-written as
\begin{equation}
\frac{T_\mathrm{mb,12}}{T_\mathrm{mb,13}} = \frac{[^{12}\mathrm{CO}]}{[^{13}\mathrm{CO}]} \frac{1-e^{-\tau_\mathrm{12}}}{\tau_\mathrm{12}},
\end{equation}
where $[^{12}\mathrm{CO}]/[^{13}\mathrm{CO}]$ is the abundance ratio, for which we use a value of 69 \citep{Wilson_1999RPPh62143W}.

We began the estimation of the correction factor \mbox{$(1-\exp{(-\tau_{12})})/\tau_{12}$} by calculating $T_\mathrm{mb,12}/T_\mathrm{mb,13}$ for each channel where both isotopologues were detected above 6$\sigma$.
In some low-velocity channels for a few FUors, the $^{13}$CO appears to be optically thick, therefore, we dropped these points from the fitting.

We then fitted a parabola 
\begin{equation}
  \frac{T_\mathrm{mb,12}}{T_\mathrm{mb,13}} = A + B\, (v-v_\mathrm{sys}) + C\,{(v-v_\mathrm{sys})}^2,
\end{equation}
which will allow us to correct for the velocity channels where the $^{13}$CO emission was not detected.
We fixed $B=0$ to keep the parabola symmetric with respect to the systemic velocity and prevent over-correcting one side of the outflow.
Finally, the correction factor selected for each channel was the lower value between the fitted parabola and the expected abundance ratio of 69.
The plots and the values of the fitted parabolas for each target are presented in \autoref{app:correction}.
We note that in the case of iPTF~15afq, due to the complex emission of $^{13}$CO, we only used blueshifted points to fit the parabola (see \autoref{app:correction}).

\begin{deluxetable*}{lccccCC}  
\tablecaption{Outflow properties from $^{12}$CO~(3--2) observations after optical depth correction.\label{tab:outflows_32_corrected}}
\tablehead{
\colhead{Target}  & \colhead{Side} & \colhead{$M_\mathrm{of}$} & \colhead{$P_\mathrm{of}$} & \colhead{$E_\mathrm{of}$} & \colhead{$F_\mathrm{of}$}  & \colhead{$L_\mathrm{of}$}\\
  & & \colhead{[M$_\odot$]} & \colhead{[M$_\odot$ \si{\kilo\meter\per\second}]} & \colhead{[erg]} & \colhead{[M$_\odot$ yr$^{-1}$ \si{\kilo\meter\per\second}]} & \colhead{[L$_\odot$]}
}
\startdata
L1551~IRS~5         & Blue & \num{2.2e-02} & \num{5.2e-02} & \num{1.4e+42} & \num{6.4e-06} & \num{1.5e-03}\\
                    & Red  & \num{3.5e-02} & \num{6.6e-02} & \num{1.4e+42} & \num{7.6e-06} & \num{1.4e-03}\\
Haro~5a~IRS         & Blue & \num{9.5e-02} & \num{2.0e-01} & \num{4.2e+42} & \num{8.2e-06} & \num{1.5e-03}\\
                    & Red  & \num{2.8e-02} & \num{8.1e-02} & \num{2.4e+42} & \num{3.7e-06} & \num{9.1e-04}\\
Reipurth~50         & Blue & \num{2.9e-02} & \num{9.2e-02} & \num{2.9e+42} & \num{4.9e-06} & \num{1.3e-03}\\
                    & Red  & \num{3.4e-02} & \num{1.6e-01} & \num{7.2e+42} & \num{5.4e-06} & \num{2.0e-03}\\
V2775~Ori           & Blue & \num{9.1e-02} & \num{2.7e-01} & \num{9.6e+42} & \num{7.8e-05} & \num{2.3e-02}\\
                    & Red  & \num{9.3e-02} & \num{2.1e-01} & \num{5.3e+42} & \num{4.8e-05} & \num{9.9e-03}\\
V899~Mon            & Blue & \num{6.8e-02} & \num{8.9e-02} & \num{1.3e+42} & \num{1.1e-06} & \num{1.3e-04}\\
                    & Red  & \num{8.5e-02} & \num{1.4e-01} & \num{2.2e+42} & \num{3.6e-06} & \num{4.9e-04}\\
V900~Mon$^\dag{}$   & Blue & \num{2.5e-02} & \num{5.1e-02} & \num{1.0e+42} & \num{1.3e-06} & \num{2.1e-04}\\
                    & Red  & \num{2.4e-01} & \num{1.8e-01} & \num{1.4e+42} & \num{1.4e-06} & \num{9.2e-05}\\
V960~Mon            & Blue & \num{1.9e-01} & \num{6.5e-01} & \num{2.4e+43} & \num{6.2e-05} & \num{1.9e-02}\\
                    & Red  & \num{8.2e-02} & \num{3.1e-01} & \num{1.2e+43} & \num{2.3e-05} & \num{7.1e-03}\\
Z~CMa               & Blue & \num{8.1e-02} & \num{1.7e-01} & \num{3.5e+42} & \num{3.8e-06} & \num{6.6e-04}\\
                    & Red  & \num{1.2e-01} & \num{3.2e-01} & \num{9.2e+42} & \num{7.8e-06} & \num{1.9e-03}\\
iPTF~15afq$^\dag{}$ & Blue & \num{2.2e-02} & \num{7.1e-02} & \num{2.3e+42} & \num{2.2e-06} & \num{6.0e-04}\\
                    & Red  & \num{3.8e-02} & \num{1.8e-01} & \num{8.9e+42} & \num{4.2e-06} & \num{1.7e-03}\\
GM~Cha              & Blue & \num{1.3e-03} & \num{2.0e-03} & \num{3.1e+40} & \num{3.1e-07} & \num{3.9e-05}\\
                    & Red  & \num{1.2e-03} & \num{3.4e-03} & \num{1.0e+41} & \num{5.4e-07} & \num{1.3e-04}\\
V346~Nor            & Blue & \num{1.9e-02} & \num{6.9e-02} & \num{2.8e+42} & \num{5.2e-06} & \num{1.8e-03}\\
                    & Red  & \num{4.6e-02} & \num{2.0e-01} & \num{9.9e+42} & \num{9.8e-06} & \num{3.9e-03}
\enddata
\tablecomments{The $\dag{}$ labels the two FUors with tentative outflow detections.}
\end{deluxetable*}

\section{Discussion}\label{sec:discussion}
Here we present our discussion about the envelope masses and their relationship with the FUor evolutionary scheme presented by \citet{Quanz_2007ApJ668359Q}.
Then we discuss the FUors for which we detected outflows, and we comment on the sources for which an outflow was not detected.
Finally, we make a statistical comparison between the properties of the outflows in our FUor sample and those from other works in the literature focused on quiescent protostars.

\subsection{Envelope masses}\label{ss:disc_envelope}
\begin{figure}
\centering
  \includegraphics[width=\linewidth]{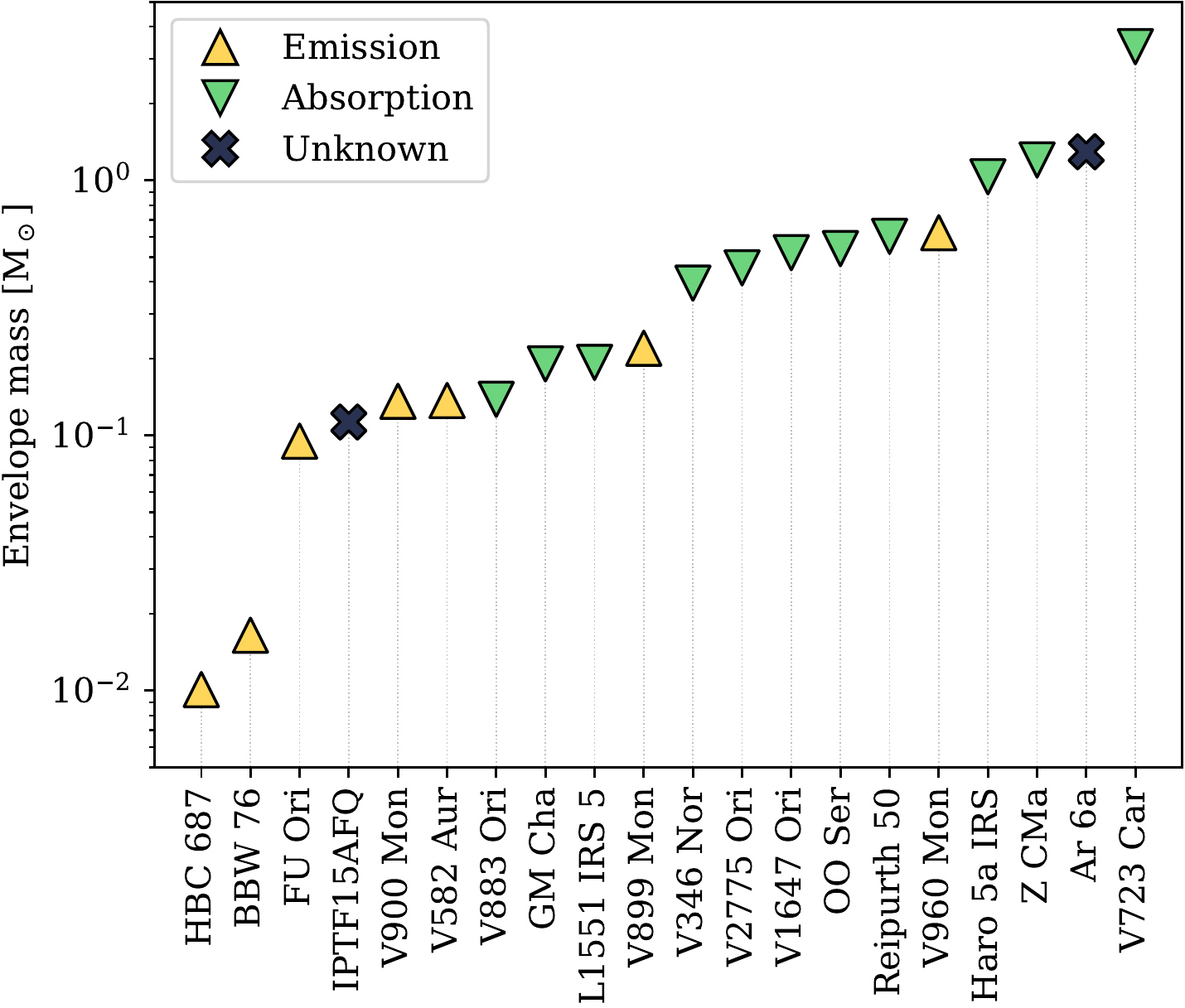}
  \caption{Comparison between envelope masses and the emission/absorption of the silicate feature.\label{fig:envelope}}
\end{figure}

\citet{Quanz_2007ApJ668359Q} targeted 14 FUor-type objects and obtained mid-infrared spectra.
They found that the silicate feature at \SI{10}{\micro\meter} could be present in either absorption or emission, and suggested that when the feature is in absorption, it is an indication of higher content of mass in the envelope surrounding the FUor and, thus, an indication of the object being younger.
In \autoref{fig:envelope} we compare our estimations of envelope masses to the emission/absorption of the silicate feature based on the references for each object listed in \autoref{tab:env_masses}.

We expanded on the work presented by \citet{Kospal_2017ApJ84345K}, who analyzed the first half of the FUor sample, and we found that the FUors with the least massive envelopes show the silicate feature in emission, while those with more massive envelopes show it in absorption.
We found two exceptions to this trend: V899~Mon and V960~Mon.
For the latter, there could be two explanations.
As mentioned below, there are three young stellar objects inside the beam of our observations, and thus, we could be significantly overestimating the amount of material in the line of sight to this FUor.
Alternatively, if we consider that the outflow is indeed driven by the FUor then based on the Moment~0 maps, we found that the direction of the outflow is aligned with the line of sight, and, therefore, it could be that the outflow has already cleared the line of sight to the FUor, allowing the detection of the silicate feature in emission while maintaining a high envelope mass.
It is harder to explain the case of V899~Mon, as our observations indicate that the direction of the outflow is perpendicular to the line of sight.
Under the assumption that the outflows are perpendicular to the inclination of the disks, we tried to verify the geometry of the systems using ALMA continuum observations \citep{Kospal_2021ApJS25630K}.
However, both disks were barely resolved and thus the inclination of uncertainties are large enough to allow the scenarios of almost edge-on and almost face-on geometries.
Observations with higher angular resolution and sensitivity are needed to determine the geometry of these systems and understand this discrepancy between envelope mass and the silicate feature.

The transition between absorption and emission appears to occur between 0.1 and 0.2\,M$_\odot$.
Indeed, V900~Mon, V582~Aur, and V883~Ori have comparable envelope masses with only the latter FUor having the silicate feature in absorption.
Here, it is not clear if the geometry of the system could explain this difference.
The inclination of V582~Aur is unknown because continuum observations have not resolved the disk \citep{Abraham_2018ApJ85328A}, and the latter two FUors have comparable inclination angles \citep{Cieza_2018MNRAS4744347C,Kospal_2021ApJS25630K}.

We do not have \SI{10}{\micro\meter} data for two sources: AR~6A and iPTF~15afq\@.
Based on its high envelope mass, we could expect the silicate feature around AR~6A to be in absorption.
However, the peak of its CO emission is off-center (\autoref{fig:moment0}) so the direct line of sight to our target could have less material and show the feature in emission.
In the case of iPTF~15afq, the peak of CO is also slightly off-center however, its mass envelope falls in the intermediate range of masses so we expect this to depend on the geometry of the system.

This suggests that HBC~687 is the most evolved FUor in our sample.
The case for the least evolved FUor is less clear as V723~Car is a massive young star and thus this evolutionary trend might not apply to it, and Z~CMa is a binary with one of its stars being an intermediate-mass star \citep{Koresko_1991AJ1022073K}.
Therefore, we consider Haro~5a~IRS as youngest FUor in our sample as it is one with the most massive envelope with the silicate feature in absorption.

\subsection{FUors with outflows}\label{ss:individual}
\paragraph{L1551~IRS~5}
This Class~I protostar was among the first detections of bipolar outflows from young stellar objects \citep{Snell_1980ApJ239L17S}.
Later observations recovered the blueshifted and redshifted lobes of the bipolar outflow in CO (2--1) \citep{MoriartySchieven_2006ApJ645357M,Wu_2009ApJ698184W} and CO (3--2) \citep{Yildiz_2015AA576A109Y}.
Based on our maps, the molecular outflows have the same geometry as seen in those previous works \citep[e.g.][]{Wu_2009ApJ698184W}, and the position angle of the outflow ($\sim$45$^\circ$) is almost perpendicular to the position angle of the circumstellar disks in the system \citep[$\sim$160$^\circ$;][]{Lim_2016ApJ826153L,FCSM_2019ApJ882L4C}.
Comparing our estimated outflow properties to those calculated by \citet{Yildiz_2015AA576A109Y}, we find that our mass estimate is in agreement with their result, while our force and luminosity are a factor of $\sim$6 lower than theirs, even when taking into account the inclination correction factor the authors applied.
However, this difference can be due to the higher sensitivity (their $v_{\max{}}$ is higher for both lobes) and the larger field of view of their observations.

\paragraph{Haro~5a~IRS}
This is a Class~I protostar is located in the Orion star forming region and it was identified as a FUor-like object by \citet{Reipurth_2012ApJ748L5R}.
Previous CO observations of source releaved its outflow \citep{Takahashi_2006ApJ651933T,Takahashi_2008ApJ688344T} with the same geometry as what we detected, including the slight overlap between the redshifted and blueshifted emission.
\citet{Kospal_2017ApJ836226K} presented the J=4--3 and J=3--2 $^{12}$CO and J=3--2 $^{13}$CO observations of this FUor, and found narrow outflow in an almost East-West direction.
Our analysis, based on the same observations as them, recovered the same morphology.
Their estimates for outflow masses are higher than ours by less than a factor of 2, which can be explained by the difference in distances and excitation temperatures, and by the difference in the velocity ranges used in the calculation of the outflow properties.
\citet{Tobin_2020ApJ890130T} and \citet{Kospal_2021ApJS25630K} presented continuum observations at millimeter wavelengths with data from ALMA and VLA, and both reported that this FUor is also a proto-binary star.

\paragraph{Reipurth~50}
A Class I protostar also referred to as HBC~494.
\citet{RuizRodriguez_2017MNRAS4663519R} presented high angular resolution observations with ALMA in which they traced the emission from the outflow in $^{12}$CO (J=2--1) and the envelope emission with the same transition of $^{13}$CO and C$^{18}$O.
The extension of the J=2--1 outflow obtained with ALMA is smaller than the size of our beam.
This can be explained by the maximum recoverable scale of their ALMA configuration (11$''$), which is comparable to the extended emission seen in their channel maps (e.g.\ the \kms{6} channel in their Figure 3), thus its is likely they have resolved out most of the extended emission of the outflow.
Indeed, most of the emission they recovered with ALMA originates from the dense cavity walls of the bipolar outflow.
Nevertheless, the position angle obtained from the high-resolution interferometric observations ($\sim$145$^\circ$) is comparable to our estimation of the position angle (150$^\circ$).
The outflow mass from the ALMA observations, calculated assuming an excitation temperature of \SI{50}{\kelvin}, is a factor of 60 higher the one we determined using the J=3--2 transition.
If we adjust for the higher temperature used in our calculations (see \autoref{ss:comparison_quiescent}), the mass estimated from ALMA measurements is still a factor of 50 higher.
This difference hints that most of the mass of the outflow of Reipurth~50 is located in the narrow cavity walls, which are severely diluted by our single-dish beam.

\paragraph{V2775~Ori}
The first detection of a molecular outflow on this object was done in the J=2--1 transitions of $^{12}$CO, $^{13}$CO and C$^{18}$O with ALMA \citep{Zurlo_2017MNRAS465834Z}.
The authors found the system is almost face-on with an inclination angle of $\sim$\SI{14}{\degree}.
Our observations recovered a similar orientation of the outflow (see \autoref{app:channelmaps}).
In addition, we found significant extended emission at both the systemic velocity and at redshifted velocities ($+$\kms{3}, see also \autoref{fig:spectra}). 
\citet{Zurlo_2017MNRAS465834Z} reported different velocity ranges for $^{12}$CO and C$^{18}$O (see their Table 2).
The velocities of the $^{12}$CO match those of the redshifted excess emission (peaking at $\sim$\kms{6}, see \autoref{fig:spectra}), and the velocities of C$^{18}$O match those of the systemic emission we report in \autoref{tab:values}.
The redshifted cloud emission appears at velocities where the outflow is still detected.
Therefore, for all the analyses of the outflow in this FUor, we removed the cloud's contamination.
Similar to the case of Reipurth~50, the difference in beam sizes and sensitivities complicate the comparison between our estimated physical properties and those from \citet{Zurlo_2017MNRAS465834Z}.
However, we found that the masses from the J=3--2 transition are higher by a factor of $\sim$8 than those estimated from the ALMA observations, which is even more surprising due to the lower excitation temperature used by \citet{Zurlo_2017MNRAS465834Z}.
We suggest that contrary on the case of Reipurth~50, the extended emission, which is likely resolved out by their interferometric observations, contains more of the mass of the outflow of V2775~Ori than the narrow cavity walls.

\paragraph{V899~Mon}
This source with a Flat or Class~II SED was originally reported as a FUor by \citet{Wils_2009ATel23071W}.
Follow-up observations indicated that the source was dimming, which was interpreted as a decrease in accretion rate by \citet{Ninan_2015ApJ8154N}.
The authors also recovered P~Cygni profile for several forbidden lines, indicating the presence of outflows.
Our observations are the first to recover an indication of a bipolar molecular outflow in CO\@, which follows a Northeast-Southwest direction.
However, the outflow position angle disagrees with the position angle of the disk \citep{Kospal_2021ApJS25630K} and of the jets detected at optical wavelengths \citep{Park_2021ApJ923171P}, thus the analysis of outflows with higher angular resolution is needed to resolve this discrepancy.
Based on its channel maps (\autoref{app:channelmaps}) there is also significant extended emission at low velocities due to the envelope.

\paragraph{V960~Mon}
Based on its pre-outburst SED, this is a Class~II object \citep{Kospal_2015ApJ801L5K} and our observations are the first to study the gas surrounding the system.
The $^{12}$CO line profiles show high-velocity wings (\autoref{fig:spectra}), which we interpreted as an indication of a bipolar molecular outflow.
The integrated emission maps (\autoref{fig:outflows} and \autoref{fig:outflows43}) and the channel maps (\autoref{app:channelmaps}) show the two outflow lobes overlapping, an indication of the outflow having a direction along the line-of-sight.
High-angular resolution ALMA observations barely resolved the FUor disk, and indicate a disk inclination between \ang{16} and \ang{60}, depending on the method used \citep{Kospal_2021ApJS25630K}.
Therefore, for the rest of the analysis, we assumed the lower inclination angle for the outflow.
\citet{Kospal_2015ApJ801L5K} detected two sources close to this FUor (one to the North and one to the Southeast), and \citet{Kospal_2021ApJS25630K} found a third one to the East.
As these sources are located within our beams, our observations contain emission from these neighboring sources, and it is possible that a source other than the FUor drives the outflow.
Therefore, our results for the outflow around this FUor must be taken with caution as an analysis of higher angular resolution observations is needed.

\paragraph{Z~CMa}
This source is a binary composed of the FUor and a Herbig~Be star with a separation of 0$\farcs$1.
\citet{Levreault_1988ApJ330897L} did not detect an outflow in the J=1--0 transition of $^{13}$CO, and in the J=2--1 and J=1--0 transitions of $^{12}$CO\@.
\citet{Evans_1994ApJ424793E} and \citet{Liljestrom_1997ApJ478381L} detected the bipolar outflow emanating from this FUor in the J=3--2 and J=1--0 transitions of CO, respectively.
Our observations recovered emission from the outflow with a Northwest-Southeast orientation (similar to that found in previous works), and we find the outflow is compact and has low velocities.
Our estimations of the outflow properties are in general agreement with those of \citet{Evans_1994ApJ424793E}.
It is unknown which of the two binary components drives the outflow, however, since both sources drive jets \citep{Whelan_2010ApJ720L119W}, it is possible that both sources drive outflows.

\paragraph{GM~Cha}
The outflows around this Class I/II object had been previously reported using a single dish antenna \citep[e.g.][]{Mottram_2017AA600A99M} and ALMA \citep{Hales_2020ApJ9007H}.
We recovered the East-West outflow orientation found by these authors.
Comparing our results with those of \citet{Mottram_2017AA600A99M}, we find the redshifted lobe is more extended than the blueshifted side.
Our estimation of mass for the blueshifted lobe is higher than their estimations, which is explained by us integrating lower velocity fluxes compared to them.
The redshifted mass and the other outflow properties are comparable to those by \citet{Mottram_2017AA600A99M}.

\paragraph{V346~Nor}
\citet{Kospal_2017ApJ836226K} presented single dish observations of the J=3--2 and J=4--3 transitions, and \citet{Kospal_2017ApJ84345K} presented ALMA Cycle~2 observations of the J=2--1 transition.
Our analysis uses the same observations as \citet{Kospal_2017ApJ836226K} and the properties of outflow have the same values within 10\%.
The small differences are due to slight differences in the methodology, such as different apertures and systemic velocities.
The orientation of the outflow is the same as that obtained at high angular resolution \citep{Kospal_2017ApJ84345K}.
Based on the $^{12}$CO/$^{13}$CO ratio used in the optical depth correction, it appears even the rarer isotopologue is optically thick at velocities close to the systemic.

\subsection{FUors with tentative detections}
Below we present the two FUors for which we can only make a tentative detection of their outflows, and thus we consider that these two sources require follow-up observations.

\paragraph{V900~Mon}
One of the most recently discovered FUors, it is a Class~I source bordering on Class~II \citep{Reipurth_2012ApJ748L5R}.
\citet{Kospal_2017ApJ84345K} used the same data as us and carried out a similar analysis as us, and did not find outflow emission.
However, \citet{Takami_2019ApJ884146T} presented high-angular resolution ALMA observations of the J=2--1 transition of $^{12}$CO, $^{13}$CO and C$^{18}$O, where they identified a bipolar outflow where the redshifted and blueshifted lobes are in the East and West directions, respectively.
Following their results, we searched for the velocity ranges that could be integrated in the J=3--2 transition for which we could find emission that following the one detected in the J=2--1 observations.
We found bipolar emission only in the J=3--2 transition that follows a similar East-West alignment (see \autoref{fig:outflows}) using the velocity range indicated in \autoref{tab:outflows_geometry}, and thus we considered this source to drive an outflow and estimated its properties.
The J=4--3 transition does not show significant emission (see \autoref{fig:outflows43}) which prompted us to consider this as only a tentative detection.

\paragraph{iPTF~15afq}
This Class~I object is one of the latest discovered FUors.
It showed a $\sim$2.5 mag brightening in 2015 which lasted a few months \citep{Miller_2015ATel74281M}, and follow-up brightenings in 2018 and 2019 \citep{Hillenbrand_2019ATel133211H}.
The 2019 outburst lasted until early 2021, and was followed by another outburst which is ongoing as of this writing\footnote{\url{http://gsaweb.ast.cam.ac.uk/alerts/alert/Gaia19fct/}}.
\citet{Hillenbrand_2019ATel133211H} presented high resolution (R=37\,000) spectra taken during outburst and found that H$\alpha$ and the Ca\,II triplet showed a P~Cygni profile, an indicator of high velocity winds.
Our observations are the first sub-millimeter wavelength observations of this object.
Its CO line profiles (\autoref{fig:spectra}) show high velocity line wings, in particular on the redshifted side.
Based on its J=3--2 integrated emission maps (\autoref{fig:outflows}) and channels maps (\autoref{app:channelmaps}), there appears to be an outflow whose blueshifted and redshifted lobes are on the Southeast and Northwest directions, respectively.
The blueshifted component is broader and with lower velocities than its redshifted counterpart.
However, we consider this FUor as only a tentative detection because the emission is heavily dominated by the envelope, and thus, it is hard to confirm that the morphology seen in the J=3--2 transition as an outflow.

\subsection{FUors without outflow detections}
ALMA observations of the J=2--1 transition showed outflow emission for two FUors: V883~Ori \citep{RuizRodriguez_2017MNRAS4683266R} and V1647~Ori \citep{Principe_2018MNRAS473879P}.
There could be multiple causes behind our lack of detection: the combination of the higher sensitivity and angular resolution in the ALMA observations, the possible low temperatures in the system, and the low velocities of the ALMA outflows.
Indeed, in the case of V883~Ori, \citet{White_2019ApJ87721W} found that the emission of $^{13}$CO J=3--2 was a combination of the outflow at low velocities and of a spherical-like envelope.
In addition, when considering interferometric observations, it is possible that they have resolved-out the contribution from the envelope, which our single-dish observations did not, therefore, the envelope emission at dominates in the low-velocity channels of our observations.
A similar case was found for FU~Ori, for which previous observations reported it did not drive an outflow \citep{Levreault_1988ApJ330897L} but ALMA observations of J=2--1 hint towards an outflow, thus making its detection uncertain \citep{Perez_2020ApJ88959P}.
We detect emission in the Northeast-Southwest direction, which is perpendicular to the position angles of the resolved disks \citep{Perez_2020ApJ88959P}; however, the angular resolution of our observations prevents us from determining if it is a bipolar outflow so we do not consider this as a detection.

For the remaining FUors (AR~6A, BBW~76, V582~Aur, V723~Car, and OO~Ser) we did not detect outflows, and we did not find previous publications that reported outflows.

\subsection{Outflow parameters}\label{ss:outflow_parameters}
We detect clear outflow emission in $\sim$55\% of the FUors in our sample (10 out of the 18), which is lower than the 92\% found in Class 0 and Class I objects \citep[e.g.][]{Mottram_2017AA600A99M}.
Even including the two cases where ALMA detected outflows when we did not (V883~Ori and V1647~Ori) and the possible outflow in FU~Ori, we would only find outflows in $\sim$73\% the FUors of our sample.
However, this is not surprising considering that some of the FUors in our sample are classified as Flat spectrum or Class~II objects, and the outflows in these evolved stages might be harder to detect due to the lower densities of the enveloping material \citep{Arce_2006ApJ6461070A}.
Indeed, only two FUors of Class~II had evidence of outflows: V960~Mon and GM~Cha.

After the optical depth correction, the outflow masses increased by a median factor of 3, with values ranging between 1 (V346~Nor) and 14 (V900~Mon).
The values of the kinematic properties (e.g.\ momentum, energy) after the optical depth correction are also a factor of a few higher than without the correction.
Even after applying this correction, we can still expect the outflow properties presented in \autoref{tab:outflows_32_corrected} to be underestimated as explained in \autoref{sss:caveats}.

\begin{figure}
\centering
\includegraphics[width=\linewidth]{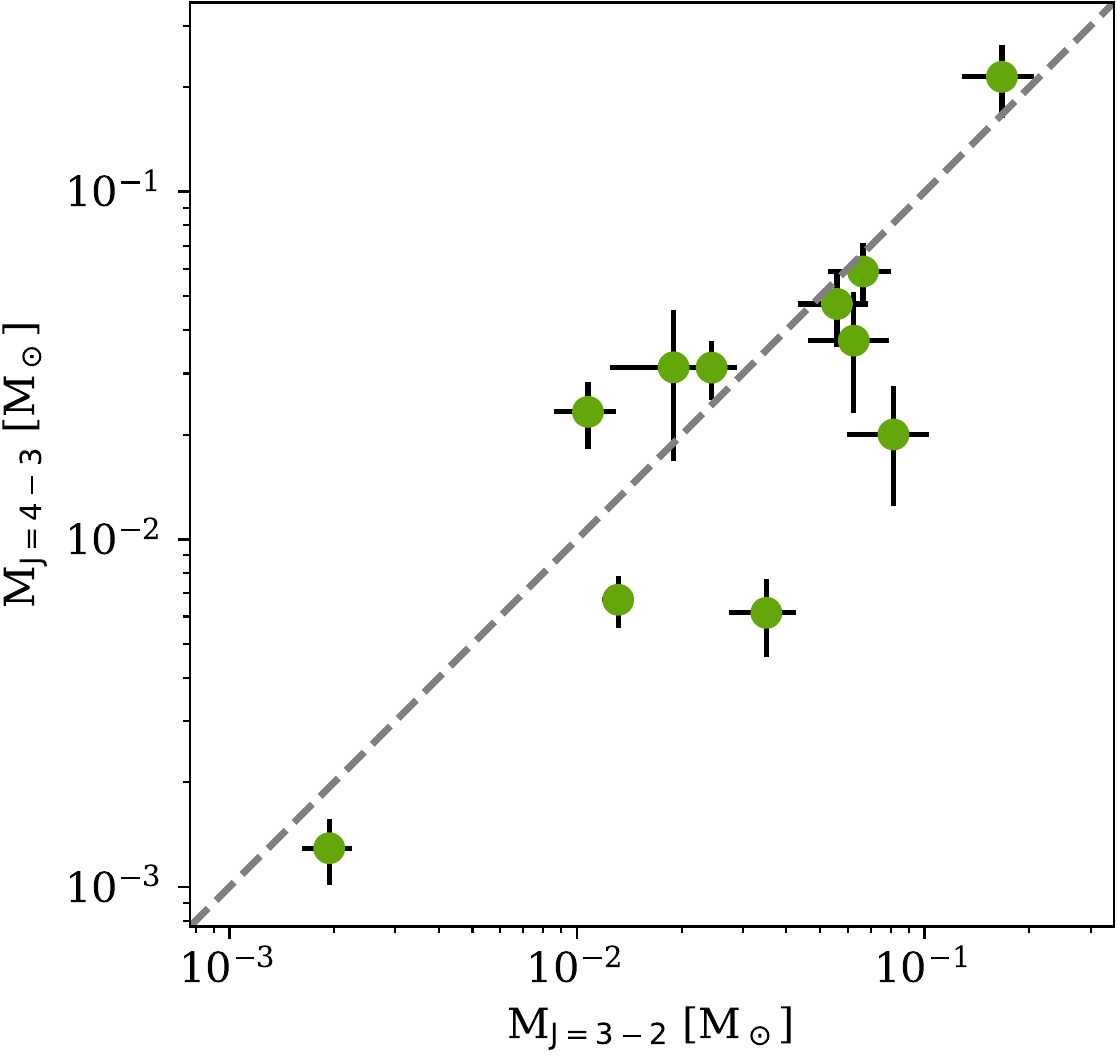}
\caption{Comparison of outflow masses determined from the J=3--2 and J=4--3 transitions. The dashed line indicates a ratio of 1.\label{fig:comparison3243}}
\end{figure}

In \autoref{fig:comparison3243} we show a comparison of the outflow masses determined for the FUors that had outflow detection in both transitions of $^{12}$CO\@.
For this comparison, we estimated the outflow mass uncertainties by multiplying the number of pixels used when calculating the mass by the rms of each data cube, and then converted these fluxes to masses using the same assumptions as the outflows.
We find that, within these uncertainties, most outflows have comparable masses in both transitions.
Reipurth~50 is the only FUor in which the mass estimate is higher in the J=4--3 transition than in J=3--2 by a factor of $\sim$2.
The outflows of V2775~Ori and V899~Mon are more massive in the lower transition by factors of 4 and 5, respectively, thus, this could be an indication of different excitation properties causing the lower transition to be stronger.
However, these comparisons are limited by differences in the observations (i.e.\ angular resolution and sensitivities), in the images (i.e.\ pixel size and field of view), and by using the same excitation for the two transitions in all the FUors.
A large-scale program to target multiple CO transitions under comparable conditions would alleviate these limitations and provide more insight on the masses of the outflows.

\subsection{Comparison with quiescent young stellar objects}\label{ss:comparison_quiescent}
We put into context our outflow properties by comparing them with the values of similar studies based on quiescent sources.
This comparison is not straightforward due to the differences in the observational properties (i.e.\ angular resolution and sensitivity), and methodology (i.e.\ choosing velocities for integration, optical depth correction and inclination correction), which have significant effects on the resulting values of the outflow properties.
It is expected that FUor outbursts last for up to a hundred years and the dynamical ages of the outflows are in the other of thousands of years (see \autoref{tab:outflows_geometry}), thus, the outflows we have detected around FUors are not related to the current outbursts.
This means that we are comparing the histories of the two samples, which could provide hints towards the nature behind the outbursts.

We compared our sample with the values of the outflow properties published in the following studies: \citet{Dunham_2014ApJ78329D}, \citet{Yildiz_2015AA576A109Y} and \citet{Mottram_2017AA600A99M}.
The three studies cover a combination of Class~0 and Class~I objects and our calculations followed similar methods to theirs.
We included quiescent Class~0 objects even when the vast majority of FUors are Class~I objects (see \autoref{tab:values}) because we want to compare how the FUor outbursts compare to the different stages of the star-formation process.
We compared the outflow masses and forces as those were the only two properties presented by all three studies from the literature.

\citet{Dunham_2014ApJ78329D} presented outflow properties with and without optical depth correction, while \citet{Yildiz_2015AA576A109Y} and \citet{Mottram_2017AA600A99M} did not calculate this correction, thus we used the optically thin values for this comparison.

\citet{Dunham_2014ApJ78329D} assumed $T_\mathrm{ex}$ = \SI{50}{\kelvin} for their calculations, while \citet{Yildiz_2015AA576A109Y}, \citet{Mottram_2017AA600A99M} used the same temperature we did in our analysis, $T_\mathrm{ex}$ = \SI{75}{\kelvin}.
To test the effect of using the lower temperature, we calculated the outflow properties of the FUors using $T_\mathrm{ex}$ = \SI{50}{\kelvin}, and found the values were a factor of $\sim$1.19 higher when using the higher temperature.
Thus, we multiplied the outflow properties of \citet{Dunham_2014ApJ78329D} by this factor to minimize the differences in methodology.
The outflow forces estimated by \citet{Yildiz_2015AA576A109Y} and \citet{Mottram_2017AA600A99M} were corrected due to the inclination of the systems based on \citet{Downes2007AA471873D}.
These correction values were estimated for Class~0 objects and are not recommended to correct Class~I objects; however, for the sake of a comparison between our sample and the ones from the literature, we applied this correction factor to the FUor outflows even if they are at later stages (Class~I or II).
We estimated the inclination of the outflows by assuming the inclination of the outflow is perpendicular to the inclination of the disk.
For most sources, we used the inclination of the FUor disks obtained from high-angular resolution observations with ALMA \citep{Cieza_2018MNRAS4744347C,Hales_2020ApJ9007H,Kospal_2021ApJS25630K}, while for Z~CMa we used the estimate by \citet{Antoniucci_2016AA593L13A} estimated from an analysis with data from optical interferometry, and in the case of iPTF~15afq, we assumed an inclination of 45$^\circ$.
However, the works studying the quiescent sample used a coarse correction table, e.g.\ Table A.5 in \citet{Mottram_2017AA600A99M}.
Thus, we rounded our inclinations to the closest values in the inclination correcion table, and the angles assumed are listed in \autoref{tab:outflows_geometry}.
We combined the quiescent samples from the literature into one, divided it by Class, and compared the two sub-samples with the FUors.

\begin{figure*}
\centering
\includegraphics[width=\textwidth]{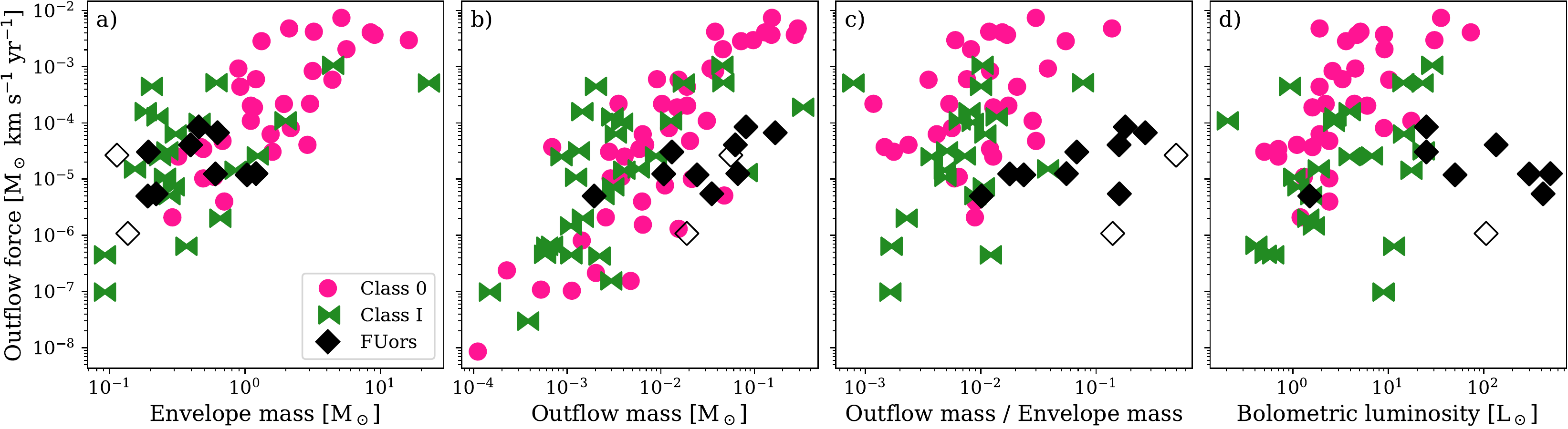}
\caption{Comparison of the outflows from quiescent sources in the literature with the FUor outflows. Outflow forces (J=3--2) plotted against envelope masses (panel a), outflow masses (J=3--2, panel b), the ratio between outflow mass and envelope mass (panel c), and bolometric luminosities (panel d). In case of the FUors, all bolometric luminosities are during outburst. The two FUors with tentative outflow detections are marked with empty diamond symbols. The $L_\mathrm{bol}$ of iPTF~15afq is unknown and thus it is not shown in panel d.}\label{fig:scattercomparison}
\end{figure*}

In \autoref{fig:scattercomparison} we plotted the outflow forces calculated from the J=3--2 transition against the envelope masses calculated from the $^{13}$CO emission (panel a), against the outflow masses also calculated from the $^{12}$CO J=3--2 transition (panel b), against the ratio between the outflow mass and envelope mass (panel c), and against the bolometric luminosities obtained from the literature (panel d; \autoref{tab:values}).

Panels a and b of \autoref{fig:scattercomparison} show that FUors outflows follow the same trends as the quiescent sources from the literature, i.e.\ higher envelope masses and higher outflow masses indicate higher outflow forces.
In the outflow mass to envelope mass ratio subplot, panel c of \autoref{fig:scattercomparison}, the FUor sample is offset from the quiescent samples.
This ratio has been used to discuss the core-to-star formation efficiency in the quiescent sample \citep{Mottram_2017AA600A99M}, thus it hints that FUors are less efficient at driving mass from the envelope onto the star, and this relationship will be discussed further below.
The values for this ratio are presented in \autoref{tab:ratio}.%
\begin{deluxetable}{lcC}
\tablecaption{Ratio between outflow masses and envelope masses using the $J$=3--2 transition of the two observe CO isotopologues, and the core-t-star formation efficiency, $\epsilon$}.\label{tab:ratio}
\tablehead{\colhead{Target} & \colhead{$M_\mathrm{outflow}$/$M_\mathrm{envelope}$} & \colhead{$\epsilon$}}
\startdata
L1551~IRS~5         & 0.068 &  -2.91 \\
Haro~5a~IRS         & 0.024 &   0.50 \\
Reipurth~50         & 0.018 &   0.69 \\
V2775~Ori           & 0.178 & -19.23 \\
V899~Mon            & 0.159 &   0.05 \\
V900~Mon$^\dag{}$   & 0.140 &   0.44 \\
V960~Mon            & 0.271 &  -8.89 \\
Z~CMa               & 0.055 &   0.39 \\
iPTF~15afq$^\dag{}$ & 0.496 &  -4.73 \\
GM~Cha              & 0.010 &   0.23 \\
V346~Nor            & 0.159 &  -2.77 \\
\enddata
\tablecomments{The $\dag{}$ labels the two FUors with tentative outflow detections.}
\end{deluxetable}%

The correlation between outflow force and bolometric luminosity, panel d of \autoref{fig:scattercomparison}, has been well studied for quiescent sources \citep{CabritBertout_1992AA261274C,Bontemps_1996AA311858B,Yildiz_2015AA576A109Y,Mottram_2017AA600A99M}, and it would appear that FUors do not follow this correlation.
However, the FUor bolometric luminosities were estimated from photometry taken while in outburst, and none have sufficient pre-outburst photometric data to estimate their pre-outburst luminosities.
Even if we do not have sufficient information about the individual FUors, when in quiescence, the protostars are expected to be low-mass and low-luminosity objects (except for V723~Car), and our measured outflow parameters are consistent with this.
Thus, our results suggest that FUors, when in quiescence, produce molecular outflows with forces comparable to those from outflows in quiescent stars.

\begin{figure*}
\centering
\includegraphics[width=\textwidth]{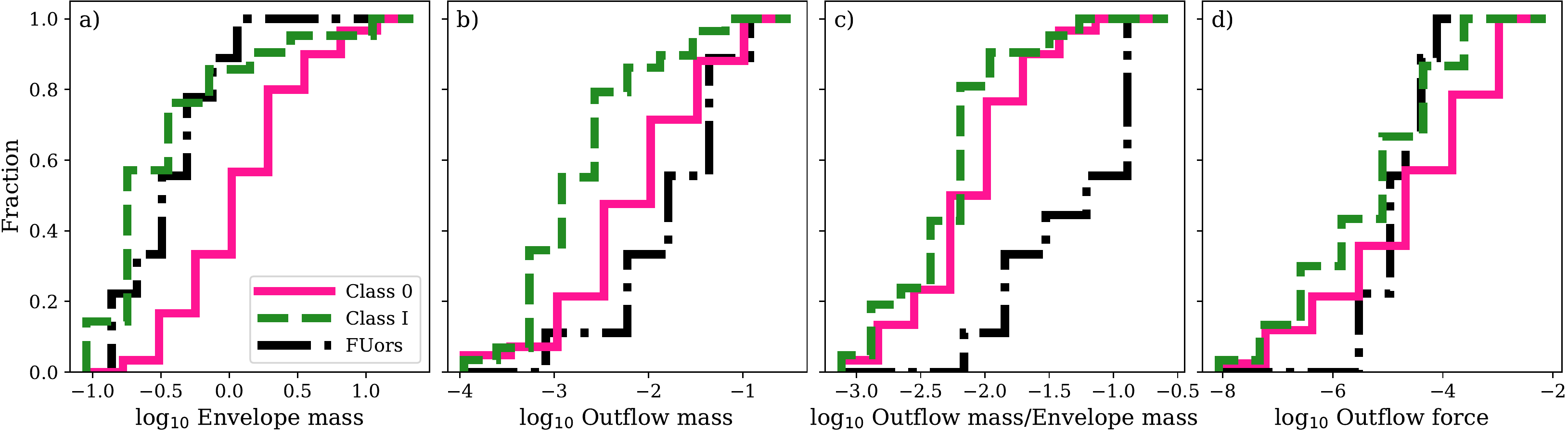}
\caption{Cumulative histograms of outflow properties, and the outflow mass to envelope mass ratio, for the Class~0 and Class~I objects from the literature \citep{Dunham_2014ApJ78329D,Yildiz_2015AA576A109Y,Mottram_2017AA600A99M}, and the FUors from this work. The bin widths for each histogram were selected using the Freedman–Diaconis rule.\label{fig:cumulativehistograms}}
\end{figure*}

In order to get a better estimate of how similar FUor outflows are with their quiescent counterparts, we present cumulative histograms comparing different properties (\autoref{fig:cumulativehistograms}), and we carried out three complementary statistical tests to examine whether the samples of the quiescent young stars were drawn from the same sample as the FUors.
The first was a two-sided Kolmogorov-Smirnov (K-S) test that compares the shapes of the distributions, the second was a Mann–Whitney U-test (MWU), which is more sensitive to the mean of the two samples rather than the shape of both distributions, and the third was a k-sample Anderson-Darling test (kAD), which is more sensitive to the tails of the distributions.
The three tests were done using the SciPy functions \texttt{kstest}, \texttt{mannwhitneyu}, and \texttt{anderson\_ksamp}, respectively.
The results of the statistical tests are presented in \autoref{tab:stats}.%

\begin{deluxetable*}{ccccccccc}
\tablecaption{$p$-values of the three statistical tests done for the envelope masses, outflow masses and forces, and the ratio between outflow and envelope masses.\label{tab:stats}}
  \tablehead{ & \multicolumn{2}{c}{Envelope masses} & \multicolumn{2}{c}{Outflow masses} & \multicolumn{2}{c}{O/E mass ratio} & \multicolumn{2}{c}{Outflow forces}\\
  \cmidrule(l{2pt}r{2pt}){2-3}  \cmidrule(l{2pt}r{2pt}){4-5} \cmidrule(l{2pt}r{2pt}){6-7} \cmidrule(l{2pt}r{2pt}){8-9}
    Test & Class~0 & Class~I & Class~0 & Class~I & Class~0 & Class~I & Class~0 & Class~I}
  \startdata
  K-S & 0.011       & 0.605    & 0.220 & \sci{1}{-3} & \sci{2}{-3} & \sci{2}{-4} & 0.070 & 0.247\\
  MWU & \sci{2}{-3} & 0.667    & 0.129 & 0.002       & \sci{5}{-4} & \sci{2}{-4} & 0.272 & 0.726\\
  kAD & $<$0.001    & $>$0.250 & 0.200 & 0.002       & $<$0.001    & $<$0.001 & 0.097 & 0.208
  \enddata
  \tablenotetext{}{The $p$-values from \texttt{anderson\_ksamp} are capped between 0.001 and 0.250, thus these values are upper or lower limits, respectively.}
\end{deluxetable*}%

With a significance level of 5\% we found that the distribution of FUor envelope masses is similar to that of Class~Is and different from the Class~0s (\autoref{fig:cumulativehistograms}, panel a), the outflow masses of FUors are different to those of Class~I objects and comparable with those of Class~0s (\autoref{fig:cumulativehistograms}, panel b), the outflow mass to envelope mass ratio is different in the FUors when compared to either sample (\autoref{fig:cumulativehistograms}, panel c), and the outflow forces of FUors are comparable with the two quiescent samples (\autoref{fig:cumulativehistograms}, panel d).
Our tests did \emph{not} include the two tentative detections. We ran the tests including these two FUors and found that our statistical results would not change.

The envelope mass result is not surprising because, based on their SEDs, the FUors are also Class~Is.
Following the same logic, the result for the outflow masses is surprising as the outflow masses are close to the Class~0s.
This could be interpreted as an indication that FUors are in the very early stages of their Class~I stage and their outflows have not had sufficient time to dissipate.
However, the masses of outflows are a combination of the material that has passed through the accretion disk and is now being driven away from the star by the outflow, and the material in the envelope that has been entrained by the outflow.
FUor outflows have higher outflow masses but similar envelope masses compared to the Class~Is, pointing towards FUor outflows having a higher percentage of material that was ejected from the accretion disk, i.e.\ material that was not accreted onto the star.
This can be seen in the ratio between the outflow mass and the envelope mass in \autoref{fig:scattercomparison} and \autoref{fig:cumulativehistograms}.

\begin{figure}
\centering
\includegraphics[width=\linewidth]{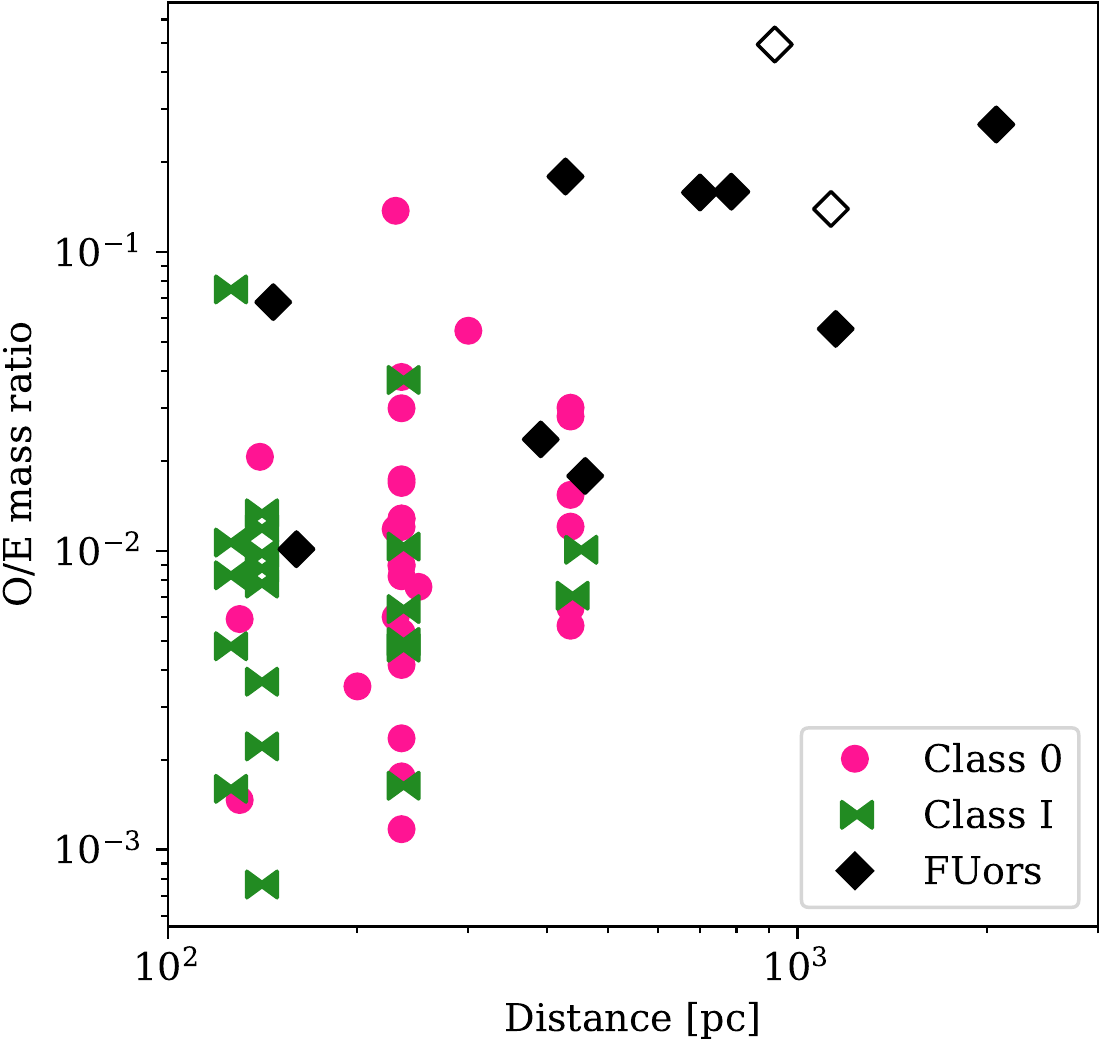}
\caption{Ratio between outflow mass and envelope mass plotted against the distance to each target. The open diamonds are the two FUors with tentative outflow detections.\label{fig:ratio_vs_distance}}
\end{figure}

The separation between the quiescent sample and the FUors might be biased due to the distance of the targets.
The YSOs of the quiescent samples are all withing 500\,pc from the Sun, while half of the FUor sample is \emph{beyond} this distance.
Therefore, if we consider that in most cases the extension of the outflows is larger than that of the envelopes, our analysis might be biased towards the FUors as it is likely that for we are measuring the full extension of their outflows in comparison to the quiescent sample.
In \autoref{fig:ratio_vs_distance} we plotted the outflow/envelope mass ratio versus the distance for each target.
The distribution of points indicates that indeed there might be a positive correlation between the mass ratio and the distance.
However, the maps in \citet{Yildiz_2015AA576A109Y} and \citet{Mottram_2017AA600A99M} indicate that 40\% of their outflows extend beyond the areas of the sky covered by their respective observations.
Therefore, the mass ratios for those sources are lower limits, and thus it raises the question whether this relationship is real or not.
An in-depth observational program covering outflows at a wide range of distances should shed some light on this matter.

\citet{Mottram_2017AA600A99M} used the outflow/envelope mass ratio to examine the core-to-star efficiency in a group of quiescent young stars.
They assumed that during the whole duration of the Class~0 and I phases, a star has an outflow rate with small enough variations that it can be approximated by a constant value, and calculate the core-to-star formation efficiency, $\epsilon$, as follows:
\begin{equation}
  \epsilon = 1 - \frac{M_{\mathrm{of}}}{M_{\mathrm{env}}} \frac{\tau_{\mathrm{0+I}}}{\tau_{\mathrm{d}}},
\end{equation}
where $\tau_{\mathrm{0+I}}$ is the total duration of the Class~0 and I phases, i.e.\ \SI{0.5}{\mega\year}.
The authors used their typical values of $M_{\mathrm{of}}/M_{\mathrm{env}}=10^{-2}$ and $\tau_{\mathrm{d}} = 10^4$ years, and found a core-to-star formation efficiency of 0.5, which is in agreement with the literature.
We used our derived masses and dynamical times to calculate $\epsilon$ for the FUors.
The values can be found in \autoref{tab:ratio}.

As can be seen from our results, five of the FUors have negative $\epsilon$ values.
These can be explained by the underestimation of the dynamical age because either the outflow appears to be face-on (V2775~Ori and V960~Mon), or it extends beyond the field of view of our observations (L1551~IRS~5, iPTF~15afq, and V346~Nor).
Three of the six FUors with positive values extend beyond our field of view (Haro~5a~IRS, Reipurth~50, and V899~Mon) and thus their $\epsilon$ values are uncertain because it is unknown how much of the mass and the extension of the outflow is beyond our field of view.
The three remaining FUors (V900~Mon, Z~CMa, and GM~Cha) with positive $\epsilon$ and with the full outflow inside the field of view, have lower efficiencies than the quiescent sample.
These values would suggest that a significant amount of material that was fed from the envelope onto the disk was not accreted onto the star but instead was driven outwards by the outflow.
However, two of these have strong caveats.
For V900~Mon, its outflow was only tentatively detected and follow-up observations might reveal a different geometry than ours, which would lead to a different value of $\epsilon$.
GM~Cha is a Class~II object, i.e.\ in a more evolved stage than most FUors, and the equation used to calculate the efficiencies was created for Class~I objects whose accretion rate is orders of magnitude higher than for Class~IIs \citep{Fiorellino_2022arXiv221107653F} and have higher envelope masses, which means that our estimated value is not accurate.

As a final point, we address the similar distributions of the outflow forces between the two quiescent samples and the FUors.
The outflow force is a property that is commonly associated to the accretion history of a young stellar object \citep{Bontemps_1996AA311858B}.
This is because, as can be seen in panel a of \autoref{fig:scattercomparison}, there is a positive correlation between the outflow force and the envelope mass, and the more evolved young stars have lower envelope masses and lower mass accretion rates.
Here we present some scenarios that can explain the lack of separation between the FUors and quiescent sample.

First, if we assume that the similar distributions are because the accretion histories of the two samples are the same then this can be interpreted as either the ``quiescent'' sample had outbursts that were undetected, or the current outbursts in the FUors are the first ones in their accretion histories.
If these are indeed the first outbursts in each of the FUors, then their effects would be undetected because the angular resolution of our observations is insufficient to resolve the inner parts of the outflows where the effects of a $<$100\,year old outburst could be detected.
Few FUor outbursts \citep[$\sim$20][]{Connelley_2018ApJ861145C} have been detected so the incidence rate of these events is unknown, and as such it is difficult to separate between these two scenarios.

Second, if we assume that the accretion histories are different between quiescent and outbursting samples then the likeness between distributions should be because of a physical property of the outflows.
The outflow force is calculated using the outflow momentum and the dynamical age of the outflow, and both of these properties depend on the distribution of velocities in the outflow.
If the FUor outflows have masses comparable to Class~0 outflows and similar velocity ranges as the outflows from the literature, we would expect the FUor outflow forces to be close to those of the Class~0 objects from the other samples.
Therefore, the divergence between Class~0 outflow forces and FUor outflow forces indicates that the latter have lower velocities.
This had already been mentioned by \citet{Principe_2018MNRAS473879P} when comparing the V1647~Ori outflow to those of others outflows observed with ALMA\@.

However, as mentioned earlier, the outflow forces presented here are highly uncertain because they are calculated as the ratio of two lower limits, the outflow momentum and the outflow dynamical age, and because of the understudied effects of the inclination of the outflow with respect to the plane of the sky.
As such, there is a need for a thorough program to study these molecular outflows.

\section{Summary \& Conclusions}\label{sec:theend}
We presented APEX observations of 20 FUors or FUor-like objects from which we estimated the envelope mass and searched for outflows.
Using a combination of line profiles and inspection of channel maps, we detected outflows in 45\% of our sample.
These include the possible first detections of molecular outflows in V899~Mon and V960~Mon, although these should be observed with higher angular resolution to corroborate them.
We also found two tentative detections in V900~Mon and iPTF~15afq, that require follow-up observations to confirm.
In the case of V883~Ori, V1647~Ori, and possibly FU~Ori, we did not detect the outflows that have been observed by ALMA\@.

Based on our $^{13}$CO measurements, envelopes with masses higher/lower than 0.1--0.2\,M$_\odot$ show the silicate feature at \SI{10}{\micro\meter} in absorption/emission.
If the envelope mass is close to this threshold level, the geometry of the system determines whether the spectral feature is in emission or absorption.
The most significant outlier of this trend is V960~Mon, which shows the \SI{10}{\micro\meter} feature in emission despite having an envelope of $\sim$0.6\,M$_\odot$.

The masses of outflows estimated from the $^{12}$CO 3--2 and 4--3 transitions are in agreement, except for two FUors: V900~Mon and V960~Mon.
We suggest that these two sources could be colder than the rest of the sample and, thus, the higher transition is dimmer.

V960~Mon is an outlier in both trends, thus we proposed another possible explanation.
This FUor has three companion YSOs in its proximity, with separations smaller than the sizes of our beams.
Therefore, we suggest that these additional sources move this object away from the trend seen in the rest of the sample.

The kinematic outflow properties (momenta, energies, forces and luminosities) are higher when estimated from the J=3--2 transition than those from J=4--3.
We attribute this to the higher sensitivity of the lower transition, which causes a difference in the range of velocities in which we detected outflow emission.

After applying an optical depth correction to the J=3--2 transition using the $^{13}$CO emission, we found that the mass of the outflows increased by a median factor of 3 and up to an order of magnitude.
The minimum improvement, seen in a few cases, showed that the outflow mass increased only by a few percent.

We compared the outflows found in our FUor sample with three works from the literature and found that outflows emanating from FUors are more massive than those from quiescent Class~I sources but with masses comparable to outflows in Class~0 sources.
We found that FUors have a higher outflow/envelope mass ratio than the quiescent sample, although this result could be biased by the distance.
We calculated the core-to-star efficiencies of the FUors and although our results are severely constrained by the geometry of the outflows, it could indicate that a significant portion of the material that was deposited into the accretion disk from the envelope is not accreted onto the star but instead is driven back to the envelope by the outflow.
Finally, we found that outflow forces from the FUor sample are comparable to the two quiescent sources, which can be interpreted as similar accretion histories or as very low velocities in the FUor outflows.

This study focused on the outflow histories of the FUors observable from the APEX site.
The dynamical ages of the detected outflows indicate that they are much older than any of the ongoing outbursts, which are less than 100 years old.
Indeed, any outflow emission directly related to the current outburst would be detected at high velocities, close to the protostar and would have small spatial scales that would be diluted by the beam of our single-dish observations.
Our comparison between the outflow properties of FUors and of other quiescent objects should be taken with caution due to the varying quality of the individual observations, and the methodology used by each research group.
A complete survey of all known FUors in both hemispheres with similar observational setups and sensitivities, and a control sample of multiple quiescent YSOs at different evolutionary stages, would greatly improve our analysis.

\begin{acknowledgments}
This project has received funding from the European Research Council (ERC) under the European Union’s Horizon 2020 research and innovation programme under grant agreement No 716155 (SACCRED).
T.Cs.\ has received financial support from the French State in the framework of the IdEx Université de Bordeaux Investments for the future Program.
This publication is based on data acquired with the Atacama Pathfinder Experiment (APEX) under programme IDs 094.F-9508 and 098.F-9505. APEX is a collaboration between the Max-Planck-Institut fur Radioastronomie, the European Southern Observatory, and the Onsala Space Observatory.
\end{acknowledgments}

\vspace{5mm}
\facilities{APEX, Herschel, JCMT}

\software{NumPy \citep{numpy}, Astropy \citep{astropy13,astropy18}, Matplotlib \citep{matplotlib}, SciPy \citep{scipy}}

\bibliography{apex_fuors.bib}

\appendix%
\restartappendixnumbering%
\section{Channel Maps}\label{app:channelmaps}
Here we present the channel maps for the three observed transitions of L1551~IRS~5, and the complete figure set with the rest of the targets in the FUor sample is available in the online journal.
The minimum and maximum velocities in the channels maps are those when the gas emission starts or finished being significant.
The purple contours are used for all the $^{13}$CO channel maps and for the $^{12}$CO maps when outflows were not detected.
The blue and red contours show the blueshifted and redshifted emission of outflows, and the green contours show the envelope emission.
All channel maps show the aperture used for the calculation of the envelope mass in the case of $^{13}$CO, and the outflow properties in the case of both $^{12}$CO transition.
The velocities shown in the plots were chosen so that the maximum and minimum velocities are shown within 27 frames, which can cause some irregular velocity steps in the plot.
However, these differences are of one channel, and we do not expect to see significant changes in the distribution of CO between two continuous channels.
\figsetstart
\figsetnum{A1}
\figsettitle{Channel maps}

\figsetgrpstart
\figsetgrpnum{A1.1}
\figsetgrptitle{V582~Aur}
\figsetplot{V582AUR_channelmaps.pdf}
\figsetgrpnote{$^{13}$CO (3--2) with contours at 3, 5, 7, 9, 11, 13, 15, 17, 19, 21 and 23$\upsigma$ with $\upsigma$ = \SI{0.36}{\kelvin}. CO (3--2) with contours at 3, 6, 9, 12, 15, 18, 21, 24, 27, 30, 33, 36, 39 and 42$\upsigma$ with $\upsigma$ = \SI{0.45}{\kelvin}. CO (4--3) with contours at 3, 5, 7, 9, 11, 13, 15, 17, 19, 21 and 23$\upsigma$ with $\upsigma$ = \SI{0.83}{\kelvin}. The black circles indicate the 10,000\,au aperture used to extract the line profiles and to calculate the systemic velocity.}
\figsetgrpend

\figsetgrpstart
\figsetgrpnum{A1.2}
\figsetgrptitle{Haro~5a~IRS}
\figsetplot{HARO5AIRS_channelmaps.pdf}
\figsetgrpnote{$^{13}$CO (3--2) with contours at 3, 5, 7, 9, 11, 13, 15, 17, 19, 21, 23, 25, 27, 29 and 31$\upsigma$ with $\upsigma$ = \SI{0.53}{\kelvin}. CO (3--2) with contours at 3, 6, 9, 12, 15, 18, 21, 24, 27, 30, 33, 36, 39 and 42$\upsigma$ with $\upsigma$ = \SI{0.56}{\kelvin}. CO (4--3) with contours at 3, 5, 7, 9, 11, 13, 15, 17, 19 and 21$\upsigma$ with $\upsigma$ = \SI{1.31}{\kelvin}. The continuous black polygon indicates the aperture used to extract the line profiles used to analyze the outflow, and the dashed circled in the $^{13}$CO channel maps indicates the 10,000\,au aperture used to calculate the systemic velocity.}
\figsetgrpend

\figsetgrpstart
\figsetgrpnum{A1.3}
\figsetgrptitle{V883~Ori}
\figsetplot{V883ORI_channelmaps.pdf}
\figsetgrpnote{$^{13}$CO (3--2) with contours at 3, 4, 5, 6, 7, 8, 9, 10, 11, 12 and 13$\upsigma$ with $\upsigma$ = \SI{0.36}{\kelvin}. CO (3--2) with contours at 3, 5, 7, 9, 11, 13, 15, 17, 19, 21, 23, 25, 27 and 29$\upsigma$ with $\upsigma$ = \SI{0.43}{\kelvin}. CO (4--3) with contours at 3, 4, 5, 6, 7, 8, 9, 10, 11, 12 and 13$\upsigma$ with $\upsigma$ = \SI{0.74}{\kelvin}. The black circles indicate the 10,000\,au aperture used to extract the line profiles and to calculate the systemic velocity.}
\figsetgrpend

\figsetgrpstart
\figsetgrpnum{A1.4}
\figsetgrptitle{Reipurth~50}
\figsetplot{REIPURTH50_channelmaps.pdf}
\figsetgrpnote{$^{13}$CO (3--2) with contours at 3, 5, 7, 9, 11, 13, 15 and 17$\upsigma$ with $\upsigma$ = \SI{0.35}{\kelvin}. CO (3--2) with contours at 3, 5, 7, 9, 11, 13, 15, 17, 19, 21 and 23$\upsigma$ with $\upsigma$ = \SI{0.46}{\kelvin}. CO (4--3) with contours at 3, 4, 5, 6, 7, 8, 9, 10, 11, 12, 13 and 14$\upsigma$ with $\upsigma$ = \SI{0.84}{\kelvin}. The continuous black polygon indicates the aperture used to extract the line profiles used to analyze the outflow, and the dashed circled in the $^{13}$CO channel maps indicates the 10,000\,au aperture used to calculate the systemic velocity.}
\figsetgrpend

\figsetgrpstart
\figsetgrpnum{A1.5}
\figsetgrptitle{FU~Ori}
\figsetplot{FUORI_channelmaps.pdf}
\figsetgrpnote{$^{13}$CO (3--2) with contours at 3, 4, 5, 6, 7, 8, 9, 10, 11, 12, 13, 14 and 15$\upsigma$ with $\upsigma$ = \SI{0.46}{\kelvin}. CO (3--2) with contours at 3, 4, 5, 6, 7, 8, 9, 10 and 11$\upsigma$ with $\upsigma$ = \SI{0.50}{\kelvin}. CO (4--3) with contours at 3, 4, 5, 6, 7, 8, 9 and 10$\upsigma$ with $\upsigma$ = \SI{0.82}{\kelvin}. The black circles indicate the 10,000\,au aperture used to extract the line profiles and to calculate the systemic velocity.}
\figsetgrpend

\figsetgrpstart
\figsetgrpnum{A1.6}
\figsetgrptitle{V1647~Ori}
\figsetplot{V1647ORI_channelmaps.pdf}
\figsetgrpnote{$^{13}$CO (3--2) with contours at 3, 5, 7, 9, 11, 13, 15, 17, 19 and 21$\upsigma$ with $\upsigma$ = \SI{0.45}{\kelvin}. CO (3--2) with contours at 3, 5, 7, 9, 11, 13, 15, 17, 19, 21, 23, 25, 27, 29 and 31$\upsigma$ with $\upsigma$ = \SI{0.48}{\kelvin}. O (4--3) with contours at 3, 4, 5, 6, 7, 8, 9, 10, 11, 12, 13, 14, 15 and 16$\upsigma$ with $\upsigma$ = \SI{0.92}{\kelvin}. The black circles indicate the 10,000\,au aperture used to extract the line profiles and to calculate the systemic velocity.}
\figsetgrpend

\figsetgrpstart
\figsetgrpnum{A1.7}
\figsetgrptitle{V2775~Ori}
\figsetplot{V2775ORI_channelmaps.pdf}
\figsetgrpnote{$^{13}$CO (3--2) with contours at 3, 4, 5, 6, 7, 8, 9, 10, 11 and 12$\upsigma$ with $\upsigma$ = \SI{0.51}{\kelvin}. CO (3--2) with contours at 3, 5, 7, 9, 11, 13, 15 and 17$\upsigma$ with $\upsigma$ = \SI{0.69}{\kelvin}. CO (4--3) with contours at 3, 4, 5, 6, 7, 8, 9, 10 and 11$\upsigma$ with $\upsigma$ = \SI{1.09}{\kelvin}. The black circles indicate the 10,000\,au aperture used to extract the line profiles to analyze the outflow and to calculate the systemic velocity.}
\figsetgrpend

\figsetgrpstart
\figsetgrpnum{A1.8}
\figsetgrptitle{V899~Mon}
\figsetplot{V899MON_channelmaps.pdf}
\figsetgrpnote{$^{13}$CO (3--2) with contours at 3, 4, 5, 6, 7, 8, 9, 10, 11, 12, 13, 14 and 15$\upsigma$ with $\upsigma$ = \SI{0.35}{\kelvin}. CO (3--2) with contours at 3, 4, 5, 6, 7, 8, 9, 10, 11, 12, 13 and 14$\upsigma$ with $\upsigma$ = \SI{0.50}{\kelvin}. CO (4--3) with contours at 3, 4, 5, 6, 7, 8, 9, 10, 11, 12, 13, 14 and 15$\upsigma$ with $\upsigma$ = \SI{0.98}{\kelvin}. The continuous black polygon indicates the aperture used to extract the line profiles used to analyze the outflow, and the dashed circled in the $^{13}$CO channel maps indicates the 10,000\,au aperture used to calculate the systemic velocity.}
\figsetgrpend

\figsetgrpstart
\figsetgrpnum{A1.9}
\figsetgrptitle{AR~6a}
\figsetplot{AR6A_channelmaps.pdf}
\figsetgrpnote{$^{13}$CO (3--2) with contours at 3, 5, 7, 9, 11, 13, 15, 17, 19, 21, 23, 25, 27 and 29$\upsigma$ with $\upsigma$ = \SI{0.48}{\kelvin}. CO (3--2) with contours at 3, 7, 11, 15, 19, 23, 27, 31, 35, 39, 43, 47 and 51$\upsigma$ with $\upsigma$ = \SI{0.59}{\kelvin}. CO (4--3) with contours at 3, 4, 5, 6, 7, 8, 9, 10, 11, 12, 13, 14 and 15$\upsigma$ with $\upsigma$ = \SI{1.77}{\kelvin}. The black circles indicate the 10,000\,au aperture used to extract the line profiles and to calculate the systemic velocity.}
\figsetgrpend

\figsetgrpstart
\figsetgrpnum{A1.10}
\figsetgrptitle{V900~Mon}
\figsetplot{V900MON_channelmaps.pdf}
\figsetgrpnote{$^{13}$CO (3--2) with contours at 3, 4, 5, 6, 7, 8, 9 and 10$\upsigma$ with $\upsigma$ = \SI{0.51}{\kelvin}. CO (3--2) with contours at 3, 4, 5, 6, 7, 8, 9, 10, 11, 12, 13, 14, 15 and 16$\upsigma$ with $\upsigma$ = \SI{0.53}{\kelvin}. CO (4--3) with contours at 3, 4, 5, 6, 7, 8 and 9$\upsigma$ with $\upsigma$ = \SI{1.31}{\kelvin}. The continuous black polygon indicates the aperture used to extract the line profiles used to analyze the outflow, and the dashed circled in the $^{13}$CO channel maps indicates the 10,000\,au aperture used to calculate the systemic velocity.}
\figsetgrpend

\figsetgrpstart
\figsetgrpnum{A1.11}
\figsetgrptitle{V960~Mon}
\figsetplot{V960MON_channelmaps.pdf}
\figsetgrpnote{$^{13}$CO (3--2) with contours at 3, 4, 5, 6, 7, 8, 9, 10, 11, 12, 13, 14 and 15$\upsigma$ with $\upsigma$ = \SI{0.35}{\kelvin}. CO (3--2) with contours at 3, 5, 7, 9, 11, 13, 15, 17, 19, 21 and 23$\upsigma$ with $\upsigma$ = \SI{0.38}{\kelvin}. CO (4--3) with contours at 3, 4, 5, 6, 7, 8, 9, 10 and 11$\upsigma$ with $\upsigma$ = \SI{0.85}{\kelvin}. The black circles indicate the 10,000\,au aperture used to extract the line profiles used to analyze the outflow and to calculate the systemic velocity.}
\figsetgrpend

\figsetgrpstart
\figsetgrpnum{A1.12}
\figsetgrptitle{Z~CMa}
\figsetplot{ZCMA_channelmaps.pdf}
\figsetgrpnote{$^{13}$CO (3--2) with contours at 3, 5, 7, 9, 11, 13, 15, 17, 19, 21, 23 and 25$\upsigma$ with $\upsigma$ = \SI{0.40}{\kelvin}. CO (3--2) with contours at 3, 6, 9, 12, 15, 18, 21, 24, 27, 30, 33, 36, 39 and 42$\upsigma$ with $\upsigma$ = \SI{0.48}{\kelvin}. CO (4--3) with contours at 3, 5, 7, 9, 11, 13, 15, 17, 19 and 21$\upsigma$ with $\upsigma$ = \SI{0.79}{\kelvin}. The continuous black polygon indicates the aperture used to extract the line profiles used to analyze the outflow, and the dashed circled in the $^{13}$CO channel maps indicates the 10,000\,au aperture used to extract the line profiles and to calculate the systemic velocity.}
\figsetgrpend

\figsetgrpstart
\figsetgrpnum{A1.13}
\figsetgrptitle{iPTF~15afq}
\figsetplot{IPTF15AFQ_channelmaps.pdf}
\figsetgrpnote{$^{13}$CO (3--2) with contours at 3, 4, 5, 6, 7, 8, 9, 10, 11, 12 and 13$\upsigma$ with $\upsigma$ = \SI{0.36}{\kelvin}. CO (3--2) with contours at 3, 4, 5, 6, 7, 8, 9, 10, 11, 12, 13, 14 and 15$\upsigma$ with $\upsigma$ = \SI{0.40}{\kelvin}. CO (4--3) with contours at 3, 4, 5, 6, 7, 8, 9, 10, 11, 12, 13, 14, 15, 16 and 17$\upsigma$ with $\upsigma$ = \SI{0.75}{\kelvin}. The continuous black polygon indicates the aperture used to extract the line profiles used to analyze the outflow, and the dashed circled in the $^{13}$CO channel maps indicates the 10,000\,au aperture used to calculate the systemic velocity.}
\figsetgrpend

\figsetgrpstart
\figsetgrpnum{A1.14}
\figsetgrptitle{BBW~76}
\figsetplot{BBW76_channelmaps.pdf}
\figsetgrpnote{$^{13}$CO (3--2) with contours at 3, 4, 5, 6, 7 and 8$\upsigma$ with $\upsigma$ = \SI{0.54}{\kelvin}. CO (3--2) with contours at 3, 4, 5, 6, 7 and 8$\upsigma$ with $\upsigma$ = \SI{0.55}{\kelvin}. CO (4--3) with contours at 3, 4, 5, 6 and 7$\upsigma$ with $\upsigma$ = \SI{1.22}{\kelvin}. The black circles indicate the 10,000\,au aperture used to extract the line profiles used to analyze the outflow and to calculate the systemic velocity.}
\figsetgrpend

\figsetgrpstart
\figsetgrpnum{A1.15}
\figsetgrptitle{V723~Car}
\figsetplot{V723CAR_channelmaps.pdf}
\figsetgrpnote{$^{13}$CO (3--2) with contours at 3, 8, 13, 18, 23, 28, 33, 38, 43, 48, 53, 58, 63, 68 and 73$\upsigma$ with $\upsigma$ = \SI{0.40}{\kelvin}. CO (3--2) with contours at 3, 12, 21, 30, 39, 48, 57, 66, 75, 84, 93, 102, 111 and 120$\upsigma$ with $\upsigma$ = \SI{0.43}{\kelvin}. CO (4--3) with contours at 3, 7, 11, 15, 19, 23, 27, 31, 35, 39, 43, 47, 51, 55 and 59$\upsigma$ with $\upsigma$ = \SI{0.92}{\kelvin}. The black circles indicate the 10,000\,au aperture used to extract the line profiles used to analyze the outflow and to calculate the systemic velocity.}
\figsetgrpend

\figsetgrpstart
\figsetgrpnum{A1.16}
\figsetgrptitle{GM~Cha}
\figsetplot{GMCHA_channelmaps.pdf}
\figsetgrpnote{$^{13}$CO (3--2) with contours at 3, 4, 5, 6, 7, 8, 9, 10, 11, 12, 13, 14, 15 and 16$\upsigma$ with $\upsigma$ = \SI{0.38}{\kelvin}. CO (3--2) with contours at 3, 5, 7, 9, 11, 13, 15, 17 and 19$\upsigma$ with $\upsigma$ = \SI{0.45}{\kelvin}. CO (4--3) with contours at 3, 4, 5, 6, 7, 8, 9, 10, 11, 12, 13, 14 and 15$\upsigma$ with $\upsigma$ = \SI{0.86}{\kelvin}. The continuous black polygon indicates the aperture used to extract the line profiles used to analyze the outflow, and the dashed circled in the $^{13}$CO channel maps indicates the 10,000\,au aperture used to calculate the systemic velocity.}
\figsetgrpend

\figsetgrpstart
\figsetgrpnum{A1.17}
\figsetgrptitle{V346~Nor}
\figsetplot{V346NOR_channelmaps.pdf}
\figsetgrpnote{$^{13}$CO (3--2) with contours at 3, 4, 5, 6, 7, 8, 9, 10, 11 and 12$\upsigma$ with $\upsigma$ = \SI{0.50}{\kelvin}. CO (3--2) with contours at 3, 5, 7, 9, 11, 13, 15, 17, 19, 21 and 23$\upsigma$ with $\upsigma$ = \SI{0.49}{\kelvin}. CO (4--3) with contours at 3, 4, 5, 6, 7, 8, 9, 10, 11, 12 and 13$\upsigma$ with $\upsigma$ = \SI{1.20}{\kelvin}. The continuous black polygon indicates the aperture used to extract the line profiles used to analyze the outflow, and the dashed circled in the $^{13}$CO channel maps indicates the 10,000\,au aperture used to calculate the systemic velocity.}
\figsetgrpend

\figsetgrpstart
\figsetgrpnum{A1.18}
\figsetgrptitle{OO~Ser}
\figsetplot{OOSER_channelmaps.pdf}
\figsetgrpnote{$^{13}$CO (3--2) with contours at 3, 4, 5, 6, 7, 8, 9, 10, 11, 12, 13, 14, 15 and 16$\upsigma$ with $\upsigma$ = \SI{0.58}{\kelvin}. CO (3--2) with contours at 3, 5, 7, 9, 11, 13, 15, 17, 19, 21, 23, 25 and 27$\upsigma$ with $\upsigma$ = \SI{0.58}{\kelvin}. CO (4--3) with contours at 3, 4, 5, 6, 7, 8, 9, 10, 11 and 12$\upsigma$ with $\upsigma$ = \SI{1.85}{\kelvin}. The black circles indicate the 10,000\,au aperture used to extract the line profiles and to calculate the systemic velocity.}
\figsetgrpend

\figsetgrpstart
\figsetgrpnum{A1.19}
\figsetgrptitle{HBC~687}
\figsetplot{HBC687_channelmaps.pdf}
\figsetgrpnote{$^{13}$CO (3--2) with contours at 3, 4, 5, 6, 7, 8 and 9$\upsigma$ with $\upsigma$ = \SI{0.61}{\kelvin}. CO (3--2) with contours at 3, 4, 5, 6 and 7$\upsigma$ with $\upsigma$ = \SI{0.63}{\kelvin}. CO (4--3) with contours at 3, 4 and 5$\upsigma$ with $\upsigma$ = \SI{1.26}{\kelvin}. The black circles indicate the 10,000\,au aperture used to extract the line profiles and to calculate the systemic velocity.}
\figsetgrpend

\figsetend

\begin{figure*}[!htb]
  \includegraphics[width=\linewidth]{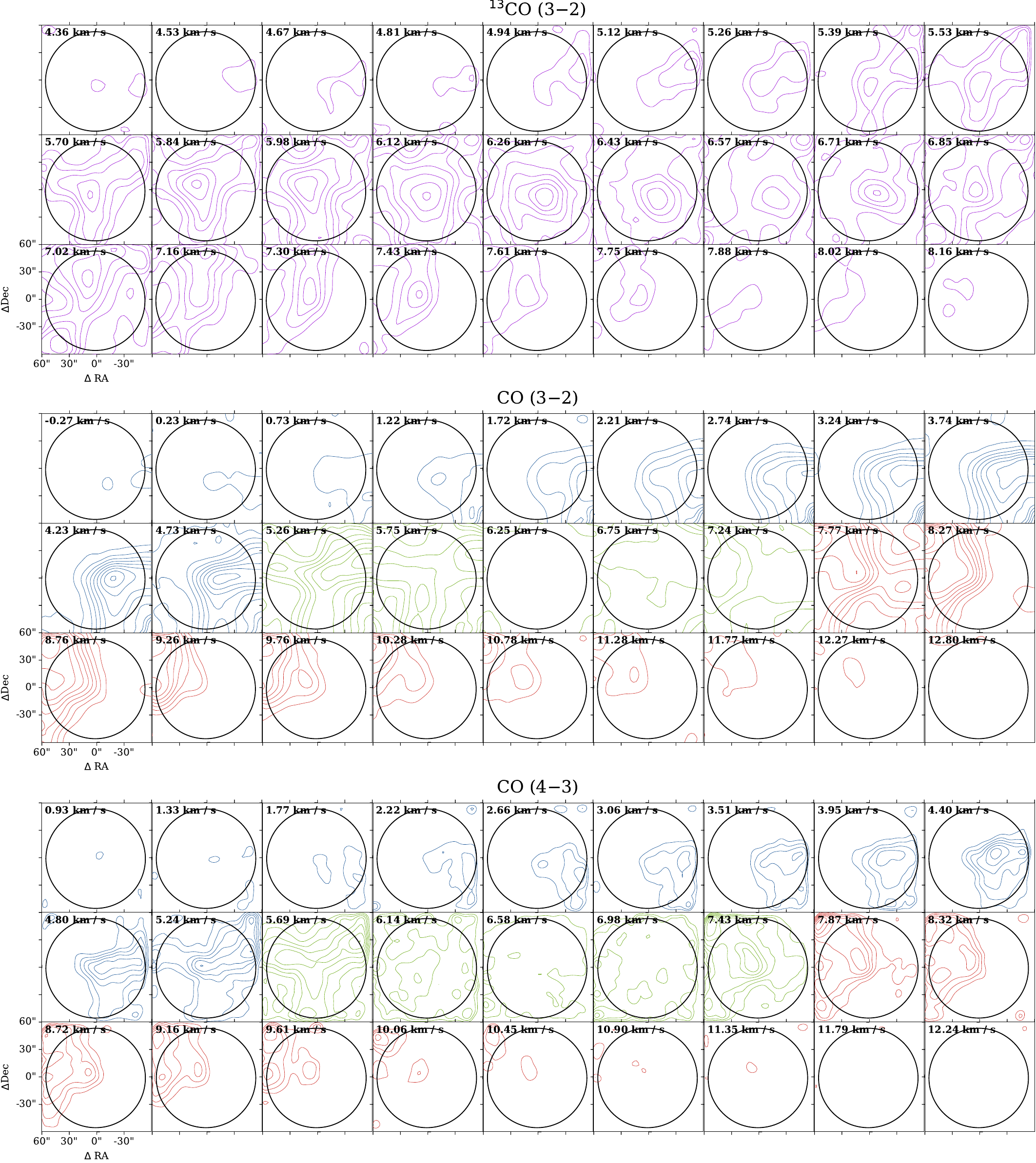}
  \caption{L1551 IRS 5 channel maps. $^{13}$CO (3--2) with contours at 3, 5, 7, 9, 11, 13, 15, 17, 19 and 21$\upsigma$ with $\upsigma$ = \SI{0.32}{\kelvin}. CO (3--2) with contours at 3, 6, 9, 12, 15, 18, 21, 24, 27, 30, 33, 36 and 39$\upsigma$ with $\upsigma$ = \SI{0.37}{\kelvin}. CO (4--3) with contours at 3, 5, 7, 9, 11, 13, 15, 17, 19 and 21$\upsigma$ with $\upsigma$ = \SI{0.65}{\kelvin}. The black circles indicate the 8000\,au aperture used to extract the line profiles used to analyze the outflow. The complete figure set (20 images) is available in the online journal.\label{fig:channelmaps_l1551irs5}}
\end{figure*}

\section{Optical Depth Correction}\label{app:correction}
\begin{figure*}[!ht]
\centering
\figurenum{B1}
\includegraphics[width=\textwidth]{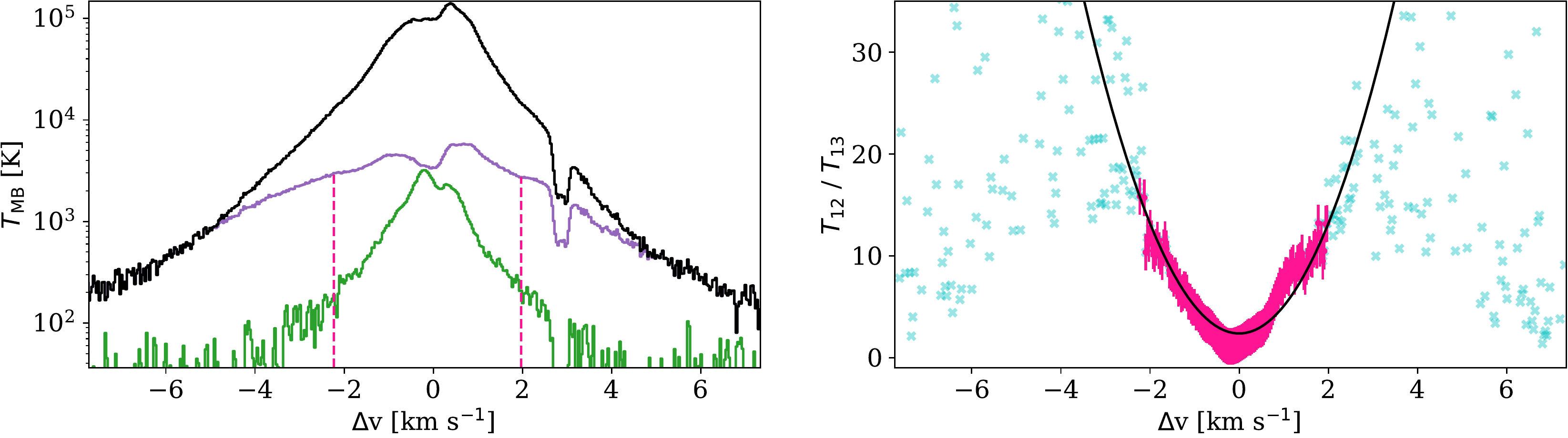}
\caption{Optical depth correction for L1551~IRS~5.
  In the left panel we show the line profiles, where the green, purple and black colors represent the $^{13}$CO, the observed $^{12}$CO and the corrected $^{12}$CO, respectively, and the vertical dashed lines indicate the rage of velocities used in the parabola fit.
  The right panel shows the ratio of main beam temperatures (T$_\mathrm{MB}$), where the light blue crosses are all the values of the ratio for each velocity channel, in pink dots with errorbars are the points used in the parabolic fitting, and the black line is the resulting best-fitted parabola. The complete figure set (10 images) is available in the online journal.\label{fig:correction_l1551irs5}}
\end{figure*}

In \autoref{tab:parabola} we present the parameters of the best-fitted parabola used to determine the optical depth correction for the sources with outflows, and in \autoref{fig:correction_l1551irs5} we present the parabolic fit and the line profiles used in the fitting.
\figsetstart
\figsetnum{B1}
\figsettitle{Optical depth correction}

\figsetgrpstart
\figsetgrpnum{B1.1}
\figsetgrptitle{Haro~5a~IRS}
\figsetplot{HARO5AIRS_parabolas.pdf}
\figsetgrpnote{Same as \autoref{fig:correction_l1551irs5} but for Haro~5a~IRS.}
\figsetgrpend

\figsetgrpstart
\figsetgrpnum{B1.2}
\figsetgrptitle{Reipurth~50}
\figsetplot{REIPURTH50_parabolas.pdf}
\figsetgrpnote{Same as \autoref{fig:correction_l1551irs5} but for Reipurth~50.}
\figsetgrpend

\figsetgrpstart
\figsetgrpnum{B1.3}
\figsetgrptitle{V2775~Ori}
\figsetplot{V2775ORI_parabolas.pdf}
\figsetgrpnote{Same as \autoref{fig:correction_l1551irs5} but for V2775~Ori.}
\figsetgrpend

\figsetgrpstart
\figsetgrpnum{B1.4}
\figsetgrptitle{V899~Mon}
\figsetplot{V899MON_parabolas.pdf}
\figsetgrpnote{Same as \autoref{fig:correction_l1551irs5} but for V899~Mon.}
\figsetgrpend

\figsetgrpstart
\figsetgrpnum{B1.5}
\figsetgrptitle{V900~Mon}
\figsetplot{V900MON_parabolas.pdf}
\figsetgrpnote{Same as \autoref{fig:correction_l1551irs5} but for V900~Mon.}
\figsetgrpend

\figsetgrpstart
\figsetgrpnum{B1.6}
\figsetgrptitle{V960~Mon}
\figsetplot{V960MON_parabolas.pdf}
\figsetgrpnote{Same as \autoref{fig:correction_l1551irs5} but for V960~Mon.}
\figsetgrpend

\figsetgrpstart
\figsetgrpnum{B1.7}
\figsetgrptitle{Z~CMa}
\figsetplot{ZCMA_parabolas.pdf}
\figsetgrpnote{Same as \autoref{fig:correction_l1551irs5} but for Z~CMa.}
\figsetgrpend

\figsetgrpstart
\figsetgrpnum{B1.8}
\figsetgrptitle{iPTF~15afq}
\figsetplot{IPTF15AFQ_parabolas.pdf}
\figsetgrpnote{Same as \autoref{fig:correction_l1551irs5} but for iPTF~15afq.}
\figsetgrpend

\figsetgrpstart
\figsetgrpnum{B1.9}
\figsetgrptitle{GM~Cha}
\figsetplot{GMCHA_parabolas.pdf}
\figsetgrpnote{Same as \autoref{fig:correction_l1551irs5} but for GM~Cha.}
\figsetgrpend

\figsetgrpstart
\figsetgrpnum{B1.10}
\figsetgrptitle{V346~Nor}
\figsetplot{V346NOR_parabolas.pdf}
\figsetgrpnote{Same as \autoref{fig:correction_l1551irs5} but for V346~Nor.}
\figsetgrpend

\figsetend

\begin{deluxetable}{ccc}[!ht]
\tablecaption{Parameters of best fitted parabolas used for optical depth correction.\label{tab:parabola}}
\tablehead{\colhead{Target} & \colhead{A} & \colhead{C}}
\startdata
L1551~IRS~5 & 2.388  & 2.683 \\
Haro~5a~IRS & 1.327  & 2.194 \\
Reipurth~50 & 2.548  & 0.643 \\
V2775~Ori   & 1.825  & 4.320 \\
V899~Mon    & 1.138  & 6.189 \\
V900~Mon    & 1.531  & 3.946 \\
V960~Mon    & 1.409  & 3.357 \\
Z~CMa       & 0.924  & 3.350 \\
iPTF~15afq  & -6.751 & 6.224 \\
GM~Cha      & -4.909 & 19.505 \\
V346~Nor    & -2.419 & 8.246 \\
\enddata
\end{deluxetable}
\end{document}